\documentclass[10pt,journal,compsoc]{IEEEtran}
\usepackage[square, comma, sort&compress, numbers]{natbib}
\usepackage[pdftex]{graphicx}   
\usepackage{amsmath}            
\usepackage{url}                
\usepackage{times}
\usepackage{courier}
\usepackage{accents}
\usepackage{amssymb}
\usepackage{amsfonts}
\usepackage{indentfirst}
\usepackage{xcolor}
\usepackage{multirow,makecell}
\usepackage{float}
\usepackage{subfigure}
\usepackage{float}
\usepackage{epstopdf}
\usepackage{bigstrut}
\usepackage{booktabs}
\usepackage{bibentry}
\usepackage{setspace}
\usepackage{CJK}
\begin{document}

\title{MV-RNN: A Multi-View Recurrent Neural Network for Sequential Recommendation}

\author{
    Qiang~Cui,
    Shu~Wu,~\IEEEmembership{Member,~IEEE,}
    Qiang~Liu,
    Wen~Zhong,
    and~Liang~Wang,~\IEEEmembership{Senior Member,~IEEE}
    \IEEEcompsocitemizethanks{
        \IEEEcompsocthanksitem
            Accepted as a regular paper of TKDE.
        \IEEEcompsocthanksitem
            Qiang Cui, Shu Wu, Qiang Liu and Liang Wang are with the Center for Research on Intelligent Perception and Computing (CRIPAC), National Laboratory of Pattern Recognition (NLPR), Institute of Automation, Chinese Academy of Sciences (CASIA) and University of Chinese Academy of Sciences (UCAS), Beijing, 100190, China.
            \protect\\ E-mail: cuiqiang2013@ia.ac.cn,
            \protect\\ \{shu.wu, qiang.liu, wangliang\}@nlpr.ia.ac.cn.
        \IEEEcompsocthanksitem
            Wen Zhong is with the University of Southern California.
            \protect\\ E-mail: wenzhong@usc.edu}}

\markboth{Journal of \LaTeX\ Class Files,~Vol.~14, No.~8, August~2015}%
{Shell \MakeLowercase{\textit{et al.}}: Bare Demo of IEEEtran.cls for Computer Society Journals}

\IEEEtitleabstractindextext{%
\begin{abstract}
    Sequential recommendation is a fundamental task for network applications, and it usually suffers from the item cold start problem due to the insufficiency of user feedbacks.
    There are currently three kinds of popular approaches which are respectively based on matrix factorization (MF) of collaborative filtering, Markov chain (MC), and recurrent neural network (RNN).
    Although widely used, they have some limitations. MF based methods could not capture dynamic user's interest. The strong Markov assumption greatly limits the performance of MC based methods.
    RNN based methods are still in the early stage of incorporating additional information. Based on these basic models, many methods with additional information only validate incorporating one modality in a separate way.
    In this work, to make the sequential recommendation and deal with the item cold start problem, we propose a \textbf{M}ulti-\textbf{V}iew \textbf{R}recurrent \textbf{N}eural \textbf{N}etwork (\textbf{MV-RNN}) model.
    Given the latent feature, MV-RNN can alleviate the item cold start problem by incorporating visual and textual information.
    First, At the input of MV-RNN, three different combinations of multi-view features are studied, like concatenation, fusion by addition and fusion by reconstructing the original multi-modal data.
    MV-RNN applies the recurrent structure to dynamically capture the user's interest. Second, we design a separate structure and a united structure on the hidden state of MV-RNN to explore a more effective way to handle multi-view features.
    Experiments on two real-world datasets show that MV-RNN can effectively generate the personalized ranking list, tackle the missing modalities problem and significantly alleviate the item cold start problem.
\end{abstract}

}

\maketitle

\IEEEdisplaynontitleabstractindextext

\IEEEpeerreviewmaketitle

\begin{figure*}[tb]
    \centering
    \setlength{\abovecaptionskip}{0pt}
    \setlength{\belowcaptionskip}{-5pt}
    \includegraphics[width=0.9\linewidth]{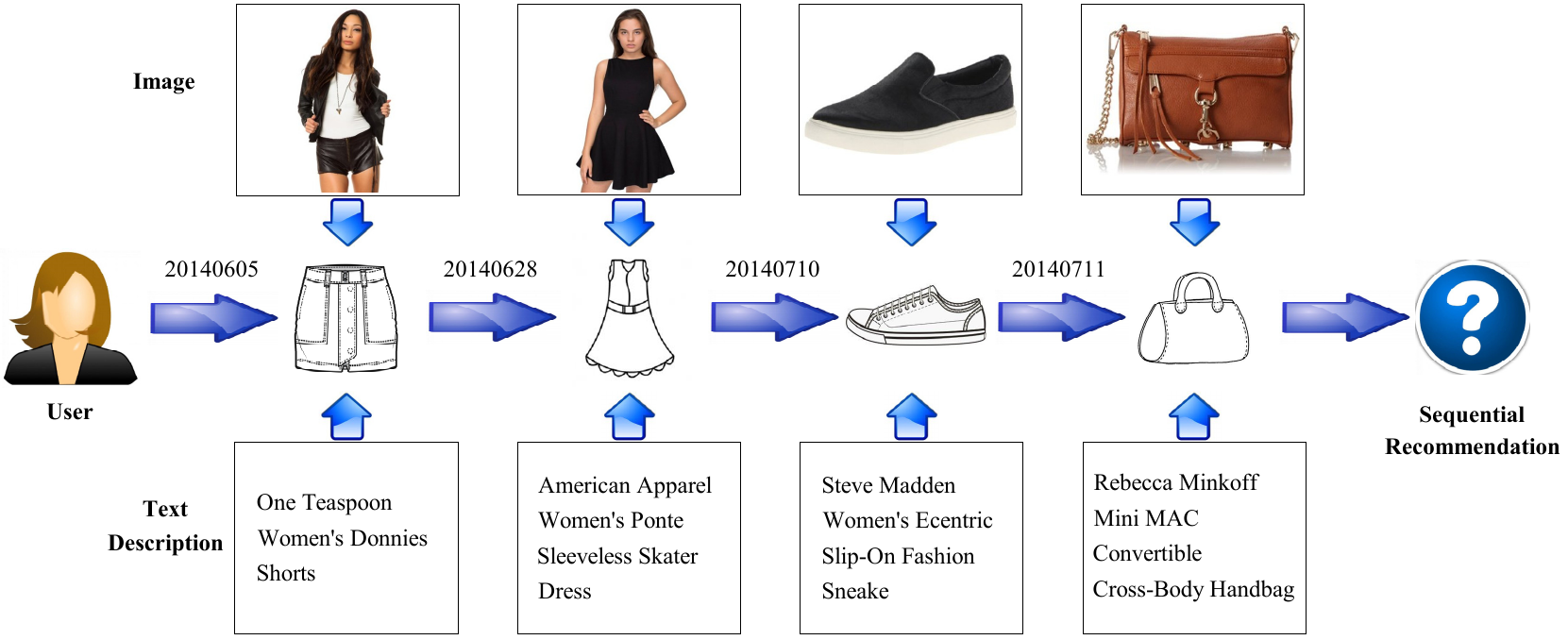}
    \caption{Diagram of a user's purchase sequence. A user buys different items at different time. We make use of the image and text description associated with each item to build the sequential recommendation model. The goal is to recommend items a user would buy in the near future, and alleviate item cold start by incorporating multiple additional information into sequence modeling.}
    \label{fig:purchase_sequence}
    \end{figure*}

\IEEEraisesectionheading{\section{Introduction}\label{sec:introduction}}
    \IEEEPARstart{R}{ecently}, with the development of Internet, applications with sequential information have become numerous and multilateral, such as web page recommendation and click prediction. Based on sequential recommendation methods, these applications could predict a user's following behaviors to improve user experience.
    Taking online shopping as an example, after a user buys an item, the application would predict a list of items that the user might buy in the near future.
    Further, we can consider the purchase behaviors as a sequence in the time order. Due to sparse user feedbacks, sequential recommendation usually encounters the item cold start problem.
    Thus, our task here concentrates on the sequential recommendation based on user historical implicit feedback and alleviating the item cold start problem.
    As shown in Figure \ref{fig:purchase_sequence}, we observe that a user will look at corresponding images and text descriptions before he or she buys items. Intuitively, we can alleviate the item cold start problem by modeling additional multi-modal information like images and text descriptions.
    Besides, we try to find a more effective way of incorporating additional information into sequence modeling.

    As for the recommendation, collaborative filtering methods are widely used. Matrix Factorization (MF) methods \cite{koren2009matrix, rendle2009bpr, salakhutdinov2011probabilistic}
    become the first choice, and learn latent representations of users and items. In order to alleviate the cold start problem, multiple additional information can be adopted, such as attribute information \cite{lu2013selective, zhao2013active}, text \cite{bao2014topicmf}, images \cite{he2016vbpr, zhao2016matrix}, and so on. Although these methods can utilize different types of features, they usually capture the user's static interest and have much difficulty in capturing sequential information. Long-term interest should be weakened while short-term interest should become prominent relatively \cite{chen2015personalized}.

    On the other hand, Markov Chain (MC) methods
    \cite{chen2015personalized, rendle2010factorizing} are widely studied for sequential recommendation by learning the transition matrix. They predict the next behavior based on recent behaviors as the transition matrix gives the probability among different states. However, MC methods could not well build the user's long-term interest due to the Markov assumption. They usually consider recent behaviors and ignore the long-term interest.
    Besides, after constructing the real world dataset of sequential scenarios like shopping and clicking, the transition probability among different states is established. The additional information no longer has any effect on this probability.

    Recently, Recurrent Neural Network (RNN) methods have shown great achievements in machine translation \cite{auli2013joint}, sequential click prediction \cite{zhang2014sequential}, location prediction \cite{liu2016predicting}, next basket recommendation \cite{yu2016dynamic}, multi-behavioral sequential prediction \cite{liu2017multi}, and so on.
    Besides, long short-term memory \cite{Hochreiter1997Long}
    and gated recurrent unit \cite{Cho2014Learning} are developed because of the gradient vanishing and explosion problem. They can hold the long-term dependency and have been applied to many tasks \cite{sutskever2014sequence, chung2015gated, hidasi2015session}.
    These RNN methods \cite{liu2016predicting, yu2016dynamic, liu2017multi} are more promising than factorizing personalized markov chains \cite{rendle2010factorizing} and other conventional MC methods.

    The existing sequential recommendation methods have difficulty in alleviating the problem of item cold start. A good choice is to apply RNN and incorporate additional multi-modal features, like images and text descriptions. Recently, the parallel RNNs model (p-RNNs) \cite{hidasi2016parallel} incorporates additional information for session-based recommendation.
    The p-RNNs model deals with multi-source data by separate subnets which are trained one by one. It builds multiple user's interests based on different views and combines the results at the end of each subset together. This way may not well leverage the advantage of multi-view data. We need to consider how to more effectively incorporate additional information to model sequential behaviors.

    In view of the above analysis, we propose a model called \textbf{M}ulti-\textbf{V}iew \textbf{R}ecurrent \textbf{N}eural \textbf{N}etwork (\textbf{MV-RNN}) for sequential recommendation and alleviating the item cold start problem.
    First, we gain visual and textual features from images and text descriptions respectively. These multi-modal features are complementary to understand the item and user's interest. A latent vector is defined for each item to represent the indirectly observable representation. These multi-view features are used as the input of MV-RNN, and three different combinations are explored. Feature concatenation and fusion naturally come to mind. More importantly, we introduce a multi-modal fusion model, called multi-modal Marginalized Denoising AutoEncoder (3mDAE). This model can help to learn more robust features and handle items with missing modalities.
    Next, we design a separate structure and a united structure for MV-RNN to explore an effective way to handle multi-view features. One applies multiple RNN units separately at every input time, and multiple hidden states of these units are concatenated together at the same time. The other employs a single RNN unit to deal with the multi-view features at once to learn a united hidden state. The MV-RNN model adopts the recurrent structure to capture dynamic changes in user's interest.
    Finally, we employ the Bayesian personalized ranking framework \cite{rendle2009bpr} and the backpropagation through time algorithm \cite{werbos1990backpropagation} to learn parameters. The main contributions are listed as follows:
        \begin{itemize}
            \item
                We design a representation of item with multi-view features. These features comprise of indirectly observable (latent) feature and directly observable (e.g., visual and textual) feature. Three combinations of multi-view features are developed, especially our 3mDAE.
            \item
                To explore a more effective way to handle multi-view inputs, MV-RNN applies a separate structure and a united structure.
                Compared to dealing with each view separately, handling multi-view features by a united structure can better leverage the advantage of different views.
            \item
                Experiments on two large real-world datasets reveal that MV-RNN is effective and outperforms the state-of-the-art methods.
        \end{itemize}

    The rest of the paper is organized as follows. Section 2 reviews previous work on sequential recommendation, cold start, and multi-modal representation learning. MV-RNN is detailly introduced in Section 3 from the perspective of input, hidden state, and output. In Section 4, we conduct extensive experiments. At last, we conclude the paper in Section 5.

\section{Related Work}
    \noindent
    In this section, we review several related works including collaborative filtering, Markov chain based methods, recurrent neural networks, and multi-modal representation learning.

\subsection{Collaborative Filtering}
    \noindent
    There are two main methods of Collaborative Filtering (CF): neighborhood models and latent factor models \cite{Koren2015Advances}. Neighborhood models have practical benefits, but they usually focus on a small subset of items or users. Latent factor models have the global perspective, and thus they tend to be more accurate. Recently, Matrix Factorization (MF) models belonging to latent factor models become fundamental because of its scalability and accuracy.
    MF absorbs rich additional information to alleviate the cold start problem, like item's attribute or user's demographics \cite{lu2013selective, zhao2013active, zhang2017enabling}. Text such as reviews is used along with the development of online searching \cite{levi2012finding}.
    Zhao et al. extend MF by combining visual data like posters and still frames of a movie to understand the movie and user's interest \cite{zhao2016matrix}. However, none of these methods could reflect the changes in user's interest over time.

    In recent years, pairwise methods become the state-of-the-art for implicit feedback \cite{he2016vbpr}. These methods can directly optimize the ranking of feedbacks and assume positive items are preferable than negative items.
    Rendle et al. \cite{rendle2009bpr} propose a Bayesian Personalized Ranking (BPR) framework to maximize the difference of user's preferences between positive and negative items. Recently, BPR is extended to combine more information like users' social relations \cite{zhao2014leveraging}. Other information like visual signals is accommodated by VBPR \cite{he2016vbpr}, which applies visual features of product images to discover user's visual interest and better understand items. Similar to MF methods, they only learn general tastes of users.


\subsection{Markov Chain Based Methods}
    \noindent
    In addition to conventional CF methods, sequential methods are popular for the recommendation and they mostly rely on Markov Chains (MC). Rendle et al. \cite{rendle2010factorizing} make a combination of MF and MC to learn both general taste and current effect for the next-basket recommendation. Chen et al. \cite{chen2015personalized} build a Markov model integrated with the forgetting mechanism to weaken long-term interest and highlight short-term interest for item recommendation.
    However, the Markov assumption hinders learning the long-term dependency because it assumes the next state only related to the last state.
    The high/variable-order MC models can make the next state related to multiple previous states, which results in a high computational cost. This problem can be solved by only considering the state-to-state probability with balancing parameters, which ignores the set-to-state probability \cite{Raftery1985A, chen2015personalized}.
    It is difficult for MC methods to model the long-term dependency.

    On the other hand, there are few Markov models involving multiple features. Chen et al. \cite{Chen2011Predictive, chen2012large} propose a two-view latent subspace Markov network to do image retrieval, annotation and so on. Their model is more like multi-view data fusion and is not suitable for sequential recommendation.
    MC is based on the probability among different states. In the sequential scenario, this probability is independent of the additional content information.

\subsection{Recurrent Neural Networks}
    \noindent
    Recently, recurrent neural networks become more and more powerful. Owing to its recurrent structure, RNN can better extract the temporal dependencies. RNN based sequential click prediction \cite{zhang2014sequential} gains the state-of-the-art performance. Yu et al. \cite{yu2016dynamic} take the representation of a basket acquired by pooling operation as the input of RNN, which is most effective for next basket recommendation. Liu et al. \cite{liu2016predicting} incorporate time-specific and distance-specific transition matrices into RNN to predict next location. Liu et al. \cite{liu2017multi} combine RNN and the Log-BiLinear model \cite{mnih2007three} to make multi-behavioral prediction. Compared with traditional sequential methods, RNN is more promising.

    Due to the gradient vanishing and explosion problem \cite{hochreiter1998vanishing, pascanu2013difficulty}, standard RNN fails to hold the long-term dependency. Lots of work have been done to alleviate this problem, and the gated activation function achieves a success, like long short-term memory (LSTM) \cite{Hochreiter1997Long} and gated recurrent unit (GRU) \cite{Cho2014Learning}.
    Sutskever et al. \cite{sutskever2014sequence} apply a multilayered LSTM to encode the input sequence and another LSTM to decode the target sequence in translation task. Their work also demonstrates LSTM can easily handle long sentences.
    Chung et al. \cite{chung2015gated} propose gated feedback RNNs to investigate the character-level language modeling. Bengio's work finds that GRU/LSTM are both certainly better than the basic RNN and GRU is comparable to LSTM on sequence modeling \cite{chung2014empirical}.

    Recently, RNN is developed to model multi-view features.
    Hidasi et al. introduce the basic RNN model to do the session-based recommendation task \cite{hidasi2015session}, then develop the p-RNNs model to incorporate rich features \cite{hidasi2016parallel}.
    The p-RNNs model builds subnets for each view separately. This is similar to the latent interest and visual interest in VBPR \cite{he2016vbpr}. Two RNNs are used to make video recommendation by using the image and make product recommendation by using text description. Compared with the basic RNN model with only ID feature, the performance improvement of p-RNNs is not significant.
    Cao et al. model multi-view features collected by the mobile phone to predict the mood score \cite{cao2017deepmood}. Obviously, there are large differences between features in their work, and they apply the late fusion to explore interactions.

\subsection{Multi-Modal Representation Learning}
    \noindent
    There are several main multi-modal representation learning methods: probabilistic graphical models, kernel-based methods and neural networks \cite{li2016multi}.
    It is often intractable and complicated to obtain exact inference for probabilistic models.
    Because of the eigenvalue problem, kernel-based methods occupy a lot of memory and time.
    On the contrary, neural networks are tractable to handle the high-dimensional data. Recently, due to the success of Deep Neural Networks (DNNs), traditional methods tend to combine deep structures.

    For methods based on DNNs, two main training strategies are widely used: Canonical Correlation Analysis (CCA) and AutoEncoder (AE) \cite{wang2015deep}. CCA based methods can make the two modalities maximally correlated. Recently, Deep CCA is proposed \cite{andrew2013deep} but it needs a large minibatch to optimize \cite{wang2015unsupervised}. Based on CCA and AE, a deep canonically correlated autoencoder model is proposed \cite{wang2015deep} for feature learning. The constraint conditions would be too complicated if CCA based methods are used in our work. Accordingly, AE based methods would be promising.

    AE based methods are very powerful to learn compact representations. AE could reproduce the input signal as far as possible and find the principal component. Vincent et al. design the denoising AE (dAE) by setting some input data to zero in a probabilistic manner \cite{Vincent2008Extracting}. After that, Vincent et al. design the stacked denoising AE (sDAE) and find that a single matrix is enough to do the encoding and decoding steps \cite{Vincent2010Stacked}. Ngiam et al. introduce the bimodal deep denoising autoencoder \cite{ngiam2011multimodal}.
    In this way, the hidden layer could learn the shared representation from different modalities. Later, Chen et al. \cite{Chen2012Marginalized} propose the marginalized denoising AE (mDAE) model, which finishes off the nonlinear transfer function and learns a linear transfer matrix. Furthermore, Wang et al. \cite{wang2016coupledmarg} propose a coupled mDAE model to deal with cross-domain learning problems. We introduce a 3mDAE model to generate multi-modal fusion representation.

\section{Proposed MV-RNN Model}
    \noindent
    In this section, we propose a Multi-View Recurrent Neural Network (MV-RNN) model. We first formulate the problem.
    Next, we explore 3 strategies to combine multi-view features at the input to represent the item.
    Then we investigate 2 structures to model multi-view features at the hidden state to build user representation.
    Finally, all the variants of MV-RNN can be trained with the Bayesian Personalized Ranking (BPR) framework and the Back Propagation Through Time (BPTT) algorithm.

    \begin{table}[tb]
      \centering\scriptsize
      \caption{Notation. }
        \begin{tabular}{ll}     
        \toprule
        Notation                & Explanation \\
        \midrule
        $\mathcal{U}$, $\mathcal{I}$, $\mathcal{I}^u$               & set of users, set of items, sequence of user $u$ \\
        $\mathcal{P}^u$, $\mathcal{V}^u$, $\mathcal{T}^u$           & sequences of training, validation and test of user $u$ \\
        $p$, $q$                & positive item, negative item \\
        $\hat{x}_{upq}^t$       & difference of preference of $u$ towards $p$ and $q$ at the $t$-th time \\
        $\mathbf{f}$, $\mathbf{g}$          & high-dimensional visual and textual features of an item \\
        $\boldsymbol{E}$, $\boldsymbol{V}$      & embedding matrices for $\mathbf{f}$, $\mathbf{g}$ \\
        $\boldsymbol{i}_\mathrm{f}$, $\boldsymbol{i}_\mathrm{g}$    & low-dimensional visual and textual features of an item \\
        $\boldsymbol{i}_\mathrm{x}$, $\boldsymbol{i}_\mathrm{m}$    & latent feature, multi-modal fusion feature built by $\boldsymbol{i}_\mathrm{f}$ and $\boldsymbol{i}_\mathrm{g}$ \\
        $d$, $d_\mathrm{f}$, $d_\mathrm{g}$         & dimensions of $\boldsymbol{i}_\mathrm{x},\mathbf{f}$, $\mathbf{g}$ \\
        $\boldsymbol{h}_\mathrm{x}$, $\boldsymbol{h}_\mathrm{m}$    & latent and multi-modal fusion features of a user \\
        $\boldsymbol{U}$, $\boldsymbol{W}$, $\boldsymbol{b}$        & transition matrices and bias for recurrent neural network \\
        \bottomrule
        \end{tabular}
      \label{table:notation}
    \end{table}

    \begin{figure*}[tb]
    \centering
    \setlength{\abovecaptionskip}{0pt}
    \setlength{\belowcaptionskip}{-5pt}
\includegraphics[width=0.9\linewidth]{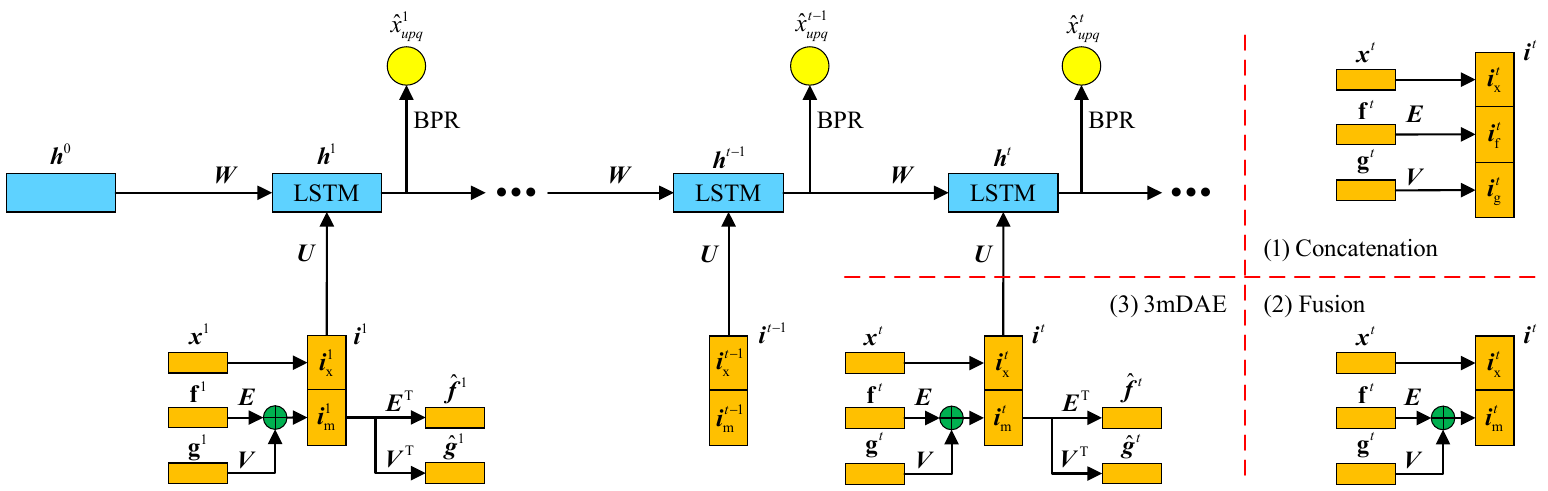}
    \caption{Diagram of the MV-RNN model. The multi-view input consists of latent feature and additional visual and textual features.
    Concatenation, Fusion and 3mDAE are three kinds of combinations of multi-view features.
    The hidden state captures dynamic changes in the user's interest.}
    \label{fig:mvgru}
    \end{figure*}

\subsection{Problem Formulation}
    \noindent
    In order to simplify the problem formulation of sequential recommendation, we take purchase histories of online shopping for instance. Let $\mathcal{U}=\{u_1,...,u_{|\mathcal{U}|}\}$ and $\mathcal{I}=\{i_1,...i_{|\mathcal{I}|}\}$ represent the sets of users and items respectively. Use $\mathcal{I}^u=(i^u_1,...,i^u_{|\mathcal{I}^u|})$ to denote the items that the user $u$ has purchased in chronological order, and the $t$-th item $i^u_t\in \mathcal{I}$. Additionally, an image and a text description are available for each item $i\in \mathcal{I}$. Given each user's history $\mathcal{I}^u$, our goal is to recommend a list of items that a user may purchase. The notation is listed in Table \ref{table:notation} for clarity.

\subsection{Representation of Item with Multi-View Features}
    \noindent
    Representation of item is used as the input of our MV-RNN model.
    Three different combinations of multi-view features are shown in Figure \ref{fig:mvgru}, and details are as follows.

    \subsubsection{Multi-View Features}
    \noindent
    There are two basic types of multi-view features of an item: indirectly observable view and directly observable view.
    The former view is latent feature, which is widely-used in recommender systems. The latent feature of an item is defined by a vector:
    \begin{equation}    \label{eq_multi_view_latent_ix}
    \boldsymbol{i}_\mathrm{x} = \boldsymbol{x},
        \qquad \qquad \qquad \boldsymbol{i}_\mathrm{x} \in \mathbb{R}^d
    \end{equation}
    The latter view refers to the additional multi-modal information that is presented externally, like image, text description, category label, video, and so on. They can provide very important information for the item. For example, image can directly show the color, text description can provide the clothing size.

    The multi-modal features consist of visual and textual features ($\mathbf{f}$ and  $\mathbf{g}$) in our work. They are obtained by GoogLeNet \cite{szegedy2015going} and GloVe \cite{pennington2014glove} weighted by TF-IDF respectively. The two kinds of features are $1024$-dimensional and $100$-dimensional vectors respectively.
    Due to the difference of $\mathbf{f}$ and  $\mathbf{g}$, we learn two linear embedding matrices $\boldsymbol{E}$ and $\boldsymbol{V}$ to transform the original high-dimensional features to embedded low-dimensional visual and textual features ($\boldsymbol{i}_\mathrm{f}$ and $\boldsymbol{i}_\mathrm{g}$):
    \begin{equation}    \label{eq_multi_view_if}
    \boldsymbol{i}_\mathrm{f} = \boldsymbol{E}\mathbf{f},
        \qquad \qquad \qquad ~ \boldsymbol{i}_\mathrm{f} \in \mathbb{R}^d
    \end{equation}
    \begin{equation}    \label{eq_multi_view_ig}
    \boldsymbol{i}_\mathrm{g} = \boldsymbol{V}\mathbf{g},
        \qquad \qquad \qquad \boldsymbol{i}_\mathrm{g} \in \mathbb{R}^d
    \end{equation}
    Sequential recommendation usually encounters the cold start problem as feedbacks are too sparse to learn fine representations of users and items. Modeling multi-view features is an effective way to alleviate this issue. These features are usually obtained from different data sources, and have different numerical ranges as well as different dimensions. Therefore, the raw features need be normalized to a same range to obtain $\boldsymbol{x}$, $\mathbf{f}$ and  $\mathbf{g}$, and should better be embedded to $d$-dimensional vectors to obtain $\boldsymbol{i}_\mathrm{x}$, $\boldsymbol{i}_\mathrm{f}$ and $\boldsymbol{i}_\mathrm{g}$. None of them is sequence data and they are aligned with each other by the item ID.

    \subsubsection{Feature Concatenation}
    \noindent
    The most natural method to combine multi-view features is concatenation. Intuitively, the item representation is $\boldsymbol{i}=\left[\boldsymbol{i}_\mathrm{x};\boldsymbol{i}_\mathrm{f};\boldsymbol{i}_\mathrm{g}\right]$. The $\boldsymbol{i}$ is a $3d$-dimensional vector, and its dimension will increase with the number of features. The capacity and complexity of this method will also increase subsequently.

    \subsubsection{Feature Fusion}
    \noindent
    Fusion can be directly established by the addition operation without nonlinear transformation:
    \begin{equation}    \label{eq_multimodal_fusion_im}
    \begin{split}
    \boldsymbol{i}_\mathrm{m} = \boldsymbol{i}_\mathrm{f} + \boldsymbol{i}_\mathrm{g},
    \qquad \qquad \boldsymbol{i}_\mathrm{m} \in \mathbb{R}^d
    \end{split}
    \end{equation}
    Please note that features with similar contents are suitable for fusion. Therefore, $\boldsymbol{i}_\mathrm{f}$ and $\boldsymbol{i}_\mathrm{g}$ are fused as the multi-modal fusion feature $\boldsymbol{i}_m$, and this process can make the model more concise. Benefiting from linear embedding and linear transformation, $\boldsymbol{i}_\mathrm{m}$ can hold all the information from $\mathbf{f}$ and $\mathbf{g}$.
    Then we obtain item representation     $\boldsymbol{i}=\left[\boldsymbol{i}_\mathrm{x};\boldsymbol{i}_\mathrm{m}\right]$ by concatenation.

    Although concatenation and fusion are easy to utilize, they still have three issues. First, both concatenation and fusion do not have an explicit objective which is able to explore correlations across modalities \cite{ngiam2011multimodal}.
    Second, they are unhandy to use in such a situation where items in the test set have missing modalities \cite{ngiam2011multimodal}. Third, no matter the combination of $\boldsymbol{i}_\mathrm{f},\boldsymbol{i}_\mathrm{g}$ is concatenation or fusion, useful information is entered into the model as well as noise. Therefore, more robust structures and parameters ($\boldsymbol{E},\boldsymbol{V}$) need to be learned.

    \begin{figure*}[tb]
    \centering
    \setlength{\abovecaptionskip}{0pt}
    \setlength{\belowcaptionskip}{-5pt}
    \subfigure[Separate Structure]{
    \begin{minipage}[b]{0.63\textwidth}
    \includegraphics[width=0.9\textwidth]{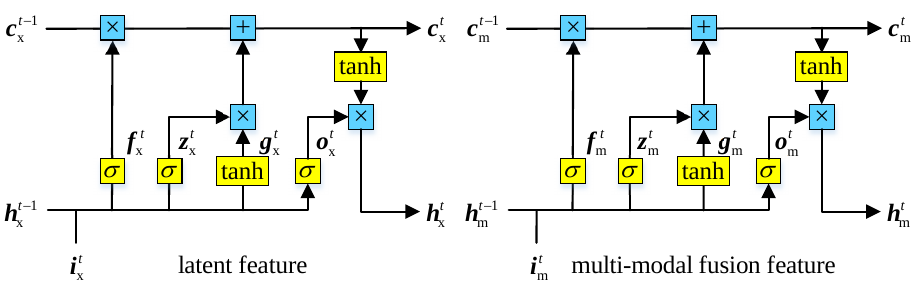}
    \label{figsub:mvgru_hidden_2units}
    \end{minipage}
    }
    \hspace{2mm}                       
    \subfigure[Unified Structure]{
    \begin{minipage}[b]{0.305\textwidth}
    \includegraphics[width=0.9\textwidth]{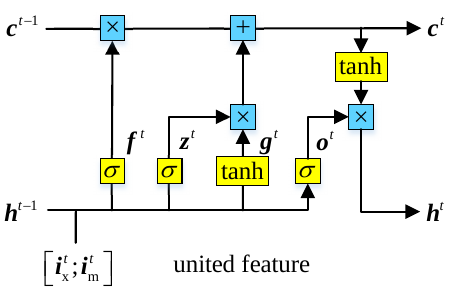}
    \label{figsub:mvgru_hidden_1unit}
    \end{minipage}
    }
    \caption{Diagram of hidden state structures of the MV-RNN model.
    We devise a separate structure and a united structure.
    The two structures handle the multi-view input features at the input by multiple RNN units and by one RNN unit each time respectively.}
    \label{fig:mvgru_hidden}
    \end{figure*}
    \subsubsection{Multi-Modal Marginalized Denoising AutoEncoder}
    \noindent
    We introduce a new fusion method to combine the multi-modal information to learn fusion feature. This method can go further to leverage the advantage of different modalities, learn more robust features and tackle the missing modalities problem.

    This method is based on the mDAE model \cite{Chen2012Marginalized}. It learns a linear mapping $\boldsymbol{M}$ and minimizes the reconstruction loss $l(\boldsymbol{t}, \boldsymbol{M}\tilde{\boldsymbol{t}})$, where $\tilde{\boldsymbol{t}}$ is the corrupted version of original feature $\boldsymbol{t}$. However, mDAE has no hidden layer.
    Later, the coupled mDAE \cite{wang2016coupledmarg} modifies the original mDAE with two mappings in a linear way $l(\boldsymbol{t}, \boldsymbol{M}^\mathrm{T}\boldsymbol{M}\tilde{\boldsymbol{t}})$. $\boldsymbol{M}\tilde{\boldsymbol{t}}$ and $\boldsymbol{M}^\mathrm{T}\boldsymbol{M}\tilde{\boldsymbol{t}}$ represent the encoding and decoding processes respectively.
    Based on these works, we introduce a \textbf{m}ulti-\textbf{m}odal \textbf{mDAE} model, called \textbf{3mDAE}, to learn fusion feature. Details are as follows.

    \noindent
    \textbf{Encoder-Decoder}.
    The encoding process is represented by Eqs. \ref{eq_multi_view_if} and \ref{eq_multi_view_ig}, and the corresponding hidden layer is built by Eq. \ref{eq_multimodal_fusion_im}. In the decoding process, we need to reconstruct the multi-modal input features. The mapping matrix in decoding process is just the transpose of the mapping matrix in encoding process \cite{Vincent2010Stacked}.
    \begin{equation}    \label{eq_decoder}
    \begin{split}
    &\hat{\boldsymbol{f}} = \boldsymbol{E}^\mathrm{T} \boldsymbol{i}_\mathrm{m} \\
    &\hat{\boldsymbol{g}} = \boldsymbol{V}^\mathrm{T} \boldsymbol{i}_\mathrm{m}
    \end{split}
    \end{equation}
    In our introduced 3mDAE model, we omit bias term and apply  original features $\mathbf{f}$ and $\mathbf{g}$ instead of corrupted version as input. The denoising operation is discussed in Section 4.3.
    The final representation of an item is also $\boldsymbol{i} = \left[ \boldsymbol{i}_\mathrm{x} ; \boldsymbol{i}_\mathrm{m} \right]$.

    \noindent
    \textbf{Objective Function.}
    The mDAE model minimizes the overall quadratic reconstruction loss for one modality \cite{wang2016coupledmarg}:
    \begin{equation}    \label{eq_optimization_mdae}    
    \Theta^* = \underaccent{\Theta}{\mathrm{argmin}}
    \frac{1}{2m}
    \sum_{i=1}^m
    \left\|
        \boldsymbol{t}_i
        - \boldsymbol{M}^\mathrm{T}\boldsymbol{M}\tilde{\boldsymbol{t}}_i
     \right\|^2,
    \end{equation}
    where $m$ is the number of samples.
    We extend this to form the objective function of 3mDAE:
    \begin{equation}    \label{eq_optimization_mmdae}    
    \Theta^* = \underaccent{\Theta}{\mathrm{argmin}}
    \frac{1}{2m} \sum_{i=1}^m
    \left(
        \frac{\| \mathbf{f}_i - \hat{\boldsymbol{f}}_i \|^2}{|d_\mathrm{f}|}
        +
        \frac{\left\| \mathbf{g}_i - \hat{\boldsymbol{g}}_i \right\|^2}{|d_\mathrm{g}|}
    \right),
    \end{equation}
    The $d_\mathrm{f}$ and $d_\mathrm{g}$ are the original dimensions of visual and textual features respectively, where $|d_\mathrm{f}|=1024$ and $|d_\mathrm{g}|=100$ in our work. They are used as balance factors.

\subsection{Modeling of Multi-View Features on Hidden State}
    \noindent
    User representation is expressed by the hidden state of our MV-RNN model. Two different ways are explored to model the multi-view features built at the input. In detail, Figures \ref{figsub:mvgru_hidden_2units} and \ref{figsub:mvgru_hidden_1unit} reveal the separate and united hidden state structures respectively. Specifically, the illustration is based on $\boldsymbol{i}_\mathrm{x}$ and $\boldsymbol{i}_\mathrm{m}$.

    \subsubsection{Long Short-Term Memory}
    \noindent
    Conventional RNN suffers from the gradient vanishing and explosion problem, so that it fails to learn long-term dependencies \cite{hochreiter1998vanishing, pascanu2013difficulty}. Gated activation function is proposed to solve this issue. We chose the widely-used LSTM \cite{Hochreiter1997Long} and it is denoted by
    \begin{equation}    \label{eq_basic_gru}
    \begin{split}
    &\boldsymbol{f}^t = \sigma \left( \boldsymbol{U}_1 \boldsymbol{x}^t + \boldsymbol{W}_1 \boldsymbol{h}^{t-1} + \boldsymbol{b}_1 \right), \\
    &\boldsymbol{z}^t = \sigma \left( \boldsymbol{U}_2 \boldsymbol{x}^t + \boldsymbol{W}_2 \boldsymbol{h}^{t-1} + \boldsymbol{b}_2 \right), \\
    &\boldsymbol{g}^t = \tanh  \left( \boldsymbol{U}_3 \boldsymbol{x}^t + \boldsymbol{W}_3 \boldsymbol{h}^{t-1} + \boldsymbol{b}_3 \right), \\
    &\boldsymbol{c}^t = \boldsymbol{f}^t \odot \boldsymbol{c}^{t-1} + \boldsymbol{z}^t \odot \boldsymbol{g}^t \\
    &\boldsymbol{o}^t = \sigma \left( \boldsymbol{U}_4 \boldsymbol{x}^t + \boldsymbol{W}_4 \boldsymbol{h}^{t-1} + \boldsymbol{b}_4 \right), \\
    &\boldsymbol{h}^t = \boldsymbol{o}^t \odot \tanh \left( \boldsymbol{c}^t \right)
    \end{split}
    \end{equation}
    where $\odot$ means element-wise product between two variables, $t$ is the time step, $\boldsymbol{x}^t \in \mathbb{R}^d$ is the input feature. Transition matrices $\boldsymbol{U}_{1\sim4} \in \mathbb{R}^{d \times d}$ transfer the current input. Recurrent connections $\boldsymbol{W}_{1\sim4} \in \mathbb{R}^{d \times d}$ delivers the sequential information. $\boldsymbol{b}_{1\sim4} \in \mathbb{R}^d$ are bias terms. The $\boldsymbol{f}^t, \boldsymbol{z}^t, \boldsymbol{g}^t, \boldsymbol{c}^t, \boldsymbol{o}^t, \boldsymbol{h}^t$ are the $forget$ gate, $input$ gate, $update$ gate, $cell$, $output$ gate and the hidden state, respectively.
    In our work, we apply a $Lstm(\cdot)$ function to substitute the original formulas in Equation \ref{eq_basic_gru}:
    \begin{equation}    \label{eq_basic_gru_unified}
    \boldsymbol{h}^t = Lstm\left( \boldsymbol{U} \boldsymbol{x}^t, \boldsymbol{W} \boldsymbol{h}^{t-1}, \boldsymbol{b} \right),
        \qquad \boldsymbol{h}^t \in \mathbb{R}^d,
    \end{equation}
    where $\boldsymbol{U}$ is a set of four matrices $\boldsymbol{U}_{1\sim4}$, and so do the $\boldsymbol{W},\boldsymbol{b}$.

    \subsubsection{Separate Multi-View RNN}
    \noindent
    A natural way to handle multi-view features is to apply separate RNN units. Each unit is used for each kind of feature. In this stage, our MV-RNN is a two-unit model, as shown in Figure \ref{figsub:mvgru_hidden_2units}.

    We apply one RNN unit to model the latent feature and apply another RNN unit to model the multi-modal fusion feature.
    Formulation is defined by:
    \begin{subequations}    \label{eq_separate_mvgru2u}
    \begin{align}
    \boldsymbol{h}_\mathrm{x}^t &= Lstm\left( \boldsymbol{U}_\mathrm{x} \boldsymbol{i}_\mathrm{x}^t, \boldsymbol{W}_\mathrm{x} \boldsymbol{h}_\mathrm{x}^{t-1}, \boldsymbol{b}_\mathrm{x} \right),
        & \boldsymbol{h}_\mathrm{x}^t &\in \mathbb{R}^d, \\
    \boldsymbol{h}_\mathrm{m}^t &= Lstm\left( \boldsymbol{U}_\mathrm{m} \boldsymbol{i}_\mathrm{m}^t, \boldsymbol{W}_\mathrm{m} \boldsymbol{h}_\mathrm{m}^{t-1}, \boldsymbol{b}_\mathrm{m} \right),
        & \boldsymbol{h}_\mathrm{m}^t &\in \mathbb{R}^d,
    \end{align}
    \end{subequations}
    where $\boldsymbol{h}_\mathrm{x}^t$ and $\boldsymbol{h}_\mathrm{m}^t$ are defined as a user's latent interest and multi-modal fusion interest at the $t$-th input.
    $\boldsymbol{U}_\mathrm{x}$ is a set of four matrices: $\boldsymbol{U}_\mathrm{x1\sim4}\in \mathbb{R}^{d \times d}$. Similarly, $\boldsymbol{W}_\mathrm{x},\boldsymbol{b}_\mathrm{x},\boldsymbol{U}_\mathrm{m},\boldsymbol{W}_\mathrm{m}$ and $\boldsymbol{b}_\mathrm{m}$ are sets of three matrices or vectors, where subscripts $\mathrm{x}$ and $\mathrm{m}$ represent the latent modeling and multi-modal modeling.


    Multi-view user representation is the concatenation of hidden states from the two RNN units. They are linked together at every time step in our work.
    \begin{equation}    \label{eq_multi-view user_representation}
    \boldsymbol{h}^t = \left[ \boldsymbol{h}_\mathrm{x}^t ; \boldsymbol{h}_\mathrm{m}^t \right],
        \qquad \boldsymbol{h}^t \in \mathbb{R}^{2d},
    \end{equation}
    where $\boldsymbol{h}^t$ is the user's general interest.
    But it may not be able to leverage the connection between multi-view features, as we model them in two RNN units separately and build discrete user's interests. Thus we tend to develop a single RNN unit to handle multi-view features simultaneously.

    \subsubsection{United Multi-View RNN}
    \noindent
    We incorporate the multi-modal fusion feature into one RNN unit together with the latent feature.
    In such situation, our MV-RNN is a one-unit model, as shown in Figure \ref{figsub:mvgru_hidden_1unit}.
    This structure can capture the relation between multi-view features and construct the united user's interest, which promotes the model to have more promising performance.
    \begin{equation}    \label{eq_unified mvgru}
    \boldsymbol{h}^t = Lstm\left( \boldsymbol{U} \left[\boldsymbol{i}_\mathrm{x}^t;\boldsymbol{i}_\mathrm{m}^t\right], \boldsymbol{W} \boldsymbol{h}^{t-1}, \boldsymbol{b} \right),
        ~ ~ ~ ~ ~ ~ \boldsymbol{h}^t \in \mathbb{R}^{2d}
    \end{equation}
    where $\boldsymbol{h}^t$ is the complete user's interest, not a simple combination of a user's different interests in Eq. \ref{eq_multi-view user_representation}.
    We apply one factor $\boldsymbol{U}$ consisting of $\boldsymbol{U}_{1\sim4}\in \mathbb{R}^{2d \times 2d}$ because we have $\left[\boldsymbol{i}_\mathrm{x}^t;\boldsymbol{i}_\mathrm{m}^t\right] \in \mathbb{R}^{2d}$, and so do the $\boldsymbol{W},\boldsymbol{b}$.

    Via the 3mDAE model and the united structure, we finally model the item's multiple (latent, visual and textual) features and the user's interest in the same feature space. Our MV-RNN model benefits from this united viewpoint.

\subsection{Model Learning}
    \noindent
    After discussing the input and hidden state of the MV-RNN model, we introduce the training procedure on output. No matter what kind of combinations of features at input or structures of hidden state, the BPR \cite{rendle2009bpr} framework is always suitable. BPR is a powerful pairwise method for implicit feedback, and it has been widely used in many works \cite{wang2015learning,hidasi2015session,liu2016predicting, yu2016dynamic,he2016vbpr,liu2017multi}. 
    Besides, as a 3mDAE model is introduced, we need to carefully consider the multi-modal reconstruction loss. A united objective function needs to be constructed. The description is also based on $\boldsymbol{i}_\mathrm{x}$ and $\boldsymbol{i}_\mathrm{m}$.

    The training set $\mathcal{S}$ is made by $(u,p,q)$ triples, where $u$ represents the user, $p$ and $q$ denote the positive and negative items respectively. Item $p$ is selected from a user's purchase history $\mathcal{I}^u$, while item $q$ is randomly chosen from the rest items ($\mathcal{I}\setminus \mathcal{I}^u$). A negative item is regenerated for each positive item in each epoch.
    \begin{equation}    \label{dataset}    
    \mathcal{S} = \left\{ {( u,p,q )} | {u \in \mathcal{U}} \wedge {p \in \mathcal{I}^u} \wedge {q \in \mathcal{I}\setminus \mathcal{I}^u} \right\}
    \end{equation}

    Given the training set, we calculate the difference of user's preferences between positive and negative items on output at every time step. At the $t$-th time step, it can be computed by
    \begin{align}   \label{preference}  
    \hat{x}_{upq}^t
    &= \hat{x}_{up}^t - \hat{x}_{uq}^t \notag \\
    &= {\left( \boldsymbol{h}^t \right)}^\mathrm{T}{\left( \boldsymbol{i}_p^{t+1} - \boldsymbol{i}_q^{t+1} \right)}
    \end{align}
    where $\boldsymbol{i}_p^{t+1}$ and $\boldsymbol{i}_q^{t+1}$ represent positive and negative inputs respectively: $\boldsymbol{i}_p^{t+1} = \left[ \boldsymbol{i}_{\mathrm{x}p}^{t+1}; \boldsymbol{i}_{\mathrm{m}p}^{t+1} \right],\boldsymbol{i}_q^{t+1} = \left[ \boldsymbol{i}_{\mathrm{x}q}^{t+1}; \boldsymbol{i}_{\mathrm{m}q}^{t+1} \right]$.

    The objective function combines BPR and our 3mDAE by a minimal form.
    The MV-RNN can simultaneously model these two kinds of losses.
    BPR maximizes the following formula:
    \begin{equation}    \label{eq_optimization_bpr}    
    \Theta^* = \underaccent{\Theta}{\mathrm{argmax}} \sum_{( u,p,q ) \in S}\ln\sigma{\left( \hat{x}_{upq} \right)} - \frac{\lambda_\Theta}{2}{\| \Theta \|^2}
    \end{equation}
    It is transformed to the minimal form in our work.
    Next, 3mDAE loss represented in Eq. \ref{eq_optimization_mmdae} is extended along with the BPR. Because we compute preference at every output using positive and negative items, we need to minimize all the visual and textual encoder-decoder losses.
    Last, we introduce a multiplicator $r_\mathrm{a}$ to leverage the preference of BPR and the reconstruction loss of our 3mDAE model. The final objective function is defined as
    \begin{equation}    \label{eq_optimization_our}    
    \begin{split}
    &\Theta^* = \underaccent{\Theta}{\mathrm{argmin}} \\
    &\sum_{( u,p,q ) \in S}
    \left(
        \begin{split}
        &- \ln\sigma{( \hat{x}_{upq} )}   \\
        &+ \frac{r_\mathrm{a}}{2|d_\mathrm{f}|}
            \left(
                \| \mathbf{f}_p - \hat{\boldsymbol{f}}_p \|^2 +
                \| \mathbf{f}_q - \hat{\boldsymbol{f}}_q \|^2
            \right) \\
        &+ \frac{r_\mathrm{a}}{2|d_\mathrm{g}|}
            \left(
                \left\| \mathbf{g}_p - \hat{\boldsymbol{g}}_p \right\|^2 +
                \left\| \mathbf{g}_q - \hat{\boldsymbol{g}}_q \right\|^2
            \right)
        \end{split}
    \right)
    + \frac{\lambda_\Theta}{2}{\| \Theta \|^2}
    \end{split}
    \end{equation}
    where $\Theta$ denotes a set of parameters $\Theta=\{\boldsymbol{X}, \boldsymbol{E}, \boldsymbol{V}, \boldsymbol{U}, \boldsymbol{W}, \boldsymbol{b} \}$. $\boldsymbol{X}$ is the set of all items' latent features. $\boldsymbol{U}$, $\boldsymbol{W}$ and $\boldsymbol{b}$ are the sets of the matrices or vectors represented in previous equations.
    $\lambda_\Theta \geqslant 0$ is the regularization parameter. Please note that $\lambda_\mathrm{ev}$ is introduced to regularize embedding matrices $\boldsymbol{E}$ and $\boldsymbol{V}$.
    Then, MV-RNN can be learned by the mini-batch gradient descent and parameters are updated by classical BPTT \cite{werbos1990backpropagation}.

    After the training, we obtain the fixed representations of $\Theta$. Then $\boldsymbol{X}, \boldsymbol{E}$ and $\boldsymbol{V}$ are reused to obtain each item's final representation. We recalculate each user's sequential hidden states, and the last hidden state denotes a user's final representation.

\section{Experimental Results and Analysis}
    \noindent
    In this section, we conduct experiments on two real-world datasets. First, experimental settings are introduced. Then a hyperparameter optimization is performed. Next, we make a comparison between MV-RNN and baselines, and a denoising experiment is conducted for our 3mDAE. The last subsection is cold start analysis on items.

    \begin{table}[tb]
      \centering\scriptsize
      \setlength{\abovecaptionskip}{0pt}
      \setlength{\belowcaptionskip}{0pt}
      \caption{Datasets. We list the numbers of users, items, feedbacks and sparsity of each dataset respectively.}
      \subtable[Datasets (5-core) used throughout the experiment.]{
        \begin{tabular}{crrrr}     
        \toprule
        dataset     & users & items & feedbacks & sparsity\\
        \midrule
        Taobao      & 1,003,331  & 343,134   & 12,613,815 & 99.996\% \\
        Amazon      & 38,840     & 22,586    & 272,949    & 99.969\% \\
        \bottomrule
        \label{tablesub:datasets_5_core}
        \end{tabular}}

      \subtable[Sub-datesets for the controlled study in Section 4.3.2.]{
        \begin{tabular}{crrrr}
        \toprule
        dataset     & users & items & feedbacks & sparsity\\
        \midrule
        Taobao (10-core)    & 478,391  & 145,867  & 7,558,233  & 99.989\% \\
        Taobao (15-core)    &  89,634  &  34,903  & 1,912,708  & 99.939\% \\
        Taobao (20-core)    &   3,536  &   1,843  &   124,453  & 98.090\% \\
        \bottomrule
        \label{tablesub:datasets_bigger_core}
        \end{tabular}}

      \label{table:datasets}
    \end{table}

    \begin{table*}[htbp]
      \centering\scriptsize
      \caption{The best parameters acquired on the validation set for all methods.}
        \begin{tabular}{cccccc|cccccc}
        \toprule
          \multirow{2}*{dataset}  &\multirow{2}*{parameter} & \multirow{2}*{BPR} &\multirow{2}*{VBPR} & \multirow{2}*{GRU/LSTM} &\multirow{2}*{p-RNN} & \multicolumn{2}{c}{based on GRU/LSTM} &\multicolumn{2}{c}{based on GRU} &\multicolumn{2}{c}{based on GRU/LSTM}\\
        \cmidrule(lr){7-8} \cmidrule(lr){9-10} \cmidrule(lr){11-12}
          & & & & & & \textbf{Con.} & \textbf{Fus.} & \textbf{3mDAE-1U} & \textbf{3mDAE-2U} & \textbf{3mDAE-1U} & \textbf{3mDAE-2U}\\

        \midrule
        \multirow{3}*{Taobao}
          & $\lambda_\Theta$        & 0.0    & 0.0    & 0.0001& 0.0001& 0.0001& 0.0001 & 0.0001 & 0.0001 & 0.0001 & 0.0001\\
          & $\lambda_\mathrm{ev}$   & -      & 0.00001& -     & -     & 0.0   & 0.0    & 0.0    & 0.0    & 0.0    & 0.0   \\
          & $r_\mathrm{a}$          & -      & -      & -     & -     & -     & -      & 0.0001 & 0.001  & 0.00001& 0.00001 \\

        \midrule
        \multirow{3}*{Amazon}
          & $\lambda_\Theta$        & 0.0001 & 0.0001 & 0.001 & 0.001 & 0.001 & 0.001  & 0.001  & 0.001  & 0.001  & 0.001 \\
          & $\lambda_\mathrm{ev}$   & -      & 0.0001 & -     & -     & 0.0   & 0.0    & 0.0    & 0.00001& 0.00001& 0.00001   \\
          & $r_\mathrm{a}$          & -      & -      & -     & -     & -     & -      & 0.0001 & 0.0001 & 0.001  & 0.001 \\
        \bottomrule
        \end{tabular}
      \label{table:parameter_lambda}
    \end{table*}

\subsection{Experimental Settings}
    \subsubsection{Datasets}
    \noindent
    Experiments are conducted on two datasets collected from Taobao\footnote{https://tianchi.shuju.aliyun.com/datalab/dataSet.htm?id=13} and Amazon\footnote{http://jmcauley.ucsd.edu/data/amazon}. The basic statistics are listed in Table \ref{table:datasets}. Both datasets have massive sequential implicit feedbacks, and each item contains an image and a text description. We apply the filtering strategy called $k$-$core$ \cite{rendle2010factorizing,wang2015learning,yu2016dynamic}. Each user purchases at least $k$ items and each item is bought by at least $k$ users. We set $k$=5 and also hold users with no more than 100 items, because users with very long sequences ($|\mathcal{I}^u|>100$) may scalp items.
    \begin{itemize}
        \item
            \textbf{Taobao} is a dataset for clothing matching competition on \emph{TianChi}\footnote{https://tianchi.shuju.aliyun.com/} platform. We use user historical data and item features (image, text) to make the sequential recommendation. Its time span is from 14-Jun-2014 to 15-Jun-2015.
        \item
            \textbf{Amazon} contains many reviews and product metadata \cite{mcauley2015image, mcauley2015inferring}. We use one large category \emph{Clothing, Shoes and Jewelry} located in the second half of the website. We acquire the sequential implicit feedback from review histories where the ratings range from 1 to 5, obtain the images and text data from product metadata. The original time span is between 29-Sep-2000 and 23-Jul-2014. As feedbacks in previous years are too sparse, we only keep feedbacks within the most recent two years.
    \end{itemize}

    \subsubsection{Multi-Modal Features}
    \noindent
    Multi-modal features are obtained by using the existing methods. They are normalized to the same range by min-max normalization. Then, they are used as the input features ($\mathbf{f}$ and  $\mathbf{g}$).

    The visual feature is obtained by the GoogLeNet \cite{szegedy2015going} implemented by BVLC Caffe deep learning framework \cite{jia2014caffe}. This network has 22 layers and has been pre-trained on 1.2M ImageNet ILSVRC2014 images \cite{russakovsky2015imagenet}. We apply the output of layer $pool5/7x7\_s1$ to obtain $1024$-dimensional visual features. They are all positive and are normalized to range $[0, 0.5]$.

    To generate the textual features of items, a text description of each item is collected firstly.
    On Taobao, we directly use item titles which have already been segmented and disordered by the data provider. On Amazon, we combine each item's category and title as its text data. Then we adopt the GloVe model \cite{pennington2014glove} weighted by TF-IDF
    \cite{salton1988term} to obtain each word's feature and weight. Finally, the weighted feature for each item is computed to obtain $100$-dimensional textual features. Their values are in the vicinity of zero and are normalized to range $[-0.5, 0.5]$.

    \subsubsection{Evaluation Metrics}
    \noindent
    Performance is evaluated on test set by Recall, Mean Average Precision (MAP) \cite{manning2008introduction} and Normalized Discounted Cumulative Gain (NDCG) \cite{wang2015learning}. The former one is an evaluation of unranked retrieval sets, while the latter two reflect the order of items. Here we consider top-$k$ (e.g., $k=$20, 30) recommendations. Besides, the Area Under the ROC Curve (AUC) \cite{rendle2009bpr,he2016vbpr} is introduced to evaluate the overall performance.

    Data is divided by time. We use feedbacks in first 60\% of the time for training, 20\% for validation and the rest 20\% for test.
    Same as p-RNNs, hyperparameters are optimized on the validation set, and all models are retrained on the full training set (training and validation sets) before obtaining final results on the test set.

    \subsubsection{Comparisons}
    \noindent
    We compare MV-RNN with several comparative baselines:
    \begin{itemize}
        \item
            \textbf{Random}: Items are randomly ranked for all users. The AUC of this method is 0.5 \cite{rendle2009bpr}.
        \item
            \textbf{POP}: This baseline recommends the most popular items in the training set for each user $u$.
        \item
            \textbf{BPR}: This method refers to the BPR-MF for implicit feedback \cite{rendle2009bpr}. It optimizes the difference of user's preferences for positive and negative items. The corresponding pairwise training procedure has been applied to many sequential tasks \cite{liu2016predicting, yu2016dynamic, liu2017multi, hidasi2015session}.
        \item
            \textbf{VBPR}: Introduced in \cite{he2016vbpr}, this is an extended method with visual features based on BPR. It firstly incorporates visual information to build the user's interest.
        \item
            \textbf{LSTM}: This sequential baseline trained with BPR is developed for next basket recommendation \cite{yu2016dynamic}. Instead of basic RNN, LSTM is used in our work. Both BPR and LSTM only model the latent feature.
        \item
            \textbf{p-RNN}: The p-RNNs is a feature-rich model for session-based recommendation \cite{hidasi2016parallel}. It has 3 structures and 4 training strategies. According to its experiments, we choose the best variant 'Parallel (res)'.
    \end{itemize}

    \begin{table*}[htbp]
      \centering\scriptsize
      \caption{The performance difference of our MV-RNN on validation set between using different baselines (GRU, LSTM). }
      \begin{tabular}{cccccc|ccccc}
        \toprule
        &\multicolumn{5}{c}{Based on GRU} &\multicolumn{5}{c}{Based on LSTM}\\
        \cmidrule(lr){2-6} \cmidrule(lr){7-11}
        \multirow{1}*{dataset} &\multirow{2}*{method}
          &\multicolumn{3}{c}{@30 (\%)} & \multirow{2}*{AUC} &\multirow{2}*{method} &\multicolumn{3}{c}{@30 (\%)} & \multirow{2}*{AUC}\\
        \cmidrule(lr){3-5} \cmidrule(lr){8-10}
           & & Recall & MAP & NDCG
           & & & Recall & MAP & NDCG\\

        \midrule
        \multirow{5}*{Taobao}
          & GRU               & 1.141 & 0.283 & 0.622 & 0.608
          & LSTM              & 1.124 & 0.287 & 0.603 & 0.610 \\
          & \textbf{Con.}     & 1.410 & 0.372 & 0.786 & 0.679
          & \textbf{Con.}     & 1.372 & 0.358 & 0.761 & 0.685 \\
          & \textbf{Fus.}     & 1.360 & 0.362 & 0.762 & 0.680
          & \textbf{Fus.}     & 1.309 & 0.332 & 0.718 & 0.678 \\
          & \textbf{3mDAE-1U} & 1.362 & 0.334 & 0.735 & 0.675
          & \textbf{3mDAE-1U} & 1.349 & 0.342 & 0.738 & 0.678 \\
          & \textbf{3mDAE-2U} & 1.186 & 0.338 & 0.690 & 0.675
          & \textbf{3mDAE-2U} & 1.196 & 0.353 & 0.709 & 0.676 \\

        \midrule
        \multirow{5}*{Amazon}
          & GRU               & 1.494 & 0.249 & 0.657 & 0.577
          & LSTM              & 1.604 & 0.305 & 0.717 & 0.583 \\
          & \textbf{Con.}     & 2.210 & 0.421 & 1.012 & 0.687
          & \textbf{Con.}     & 2.250 & 0.433 & 1.049 & 0.685 \\
          & \textbf{Fus.}     & 2.091 & 0.418 & 0.962 & 0.687
          & \textbf{Fus.}     & 2.248 & 0.415 & 0.998 & 0.687 \\
          & \textbf{3mDAE-1U} & 2.237 & 0.410 & 1.013 & 0.688
          & \textbf{3mDAE-1U} & 2.237 & 0.430 & 1.038 & 0.685 \\
          & \textbf{3mDAE-2U} & 2.104 & 0.401 & 0.955 & 0.687
          & \textbf{3mDAE-2U} & 2.283 & 0.425 & 1.035 & 0.690 \\

        \bottomrule
        \end{tabular}
      \label{table:gru_vs_lstm}
    \end{table*}

    \begin{figure*}[htbp]
    \centering
    \setlength{\abovecaptionskip}{0pt}
    \setlength{\belowcaptionskip}{-7pt}

    \includegraphics[width=0.6\textwidth]{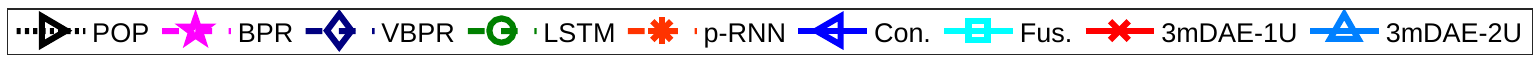}

    \subfigure[Taobao]{
    \begin{minipage}[b]{0.8\textwidth}
    \includegraphics[width=1\textwidth]{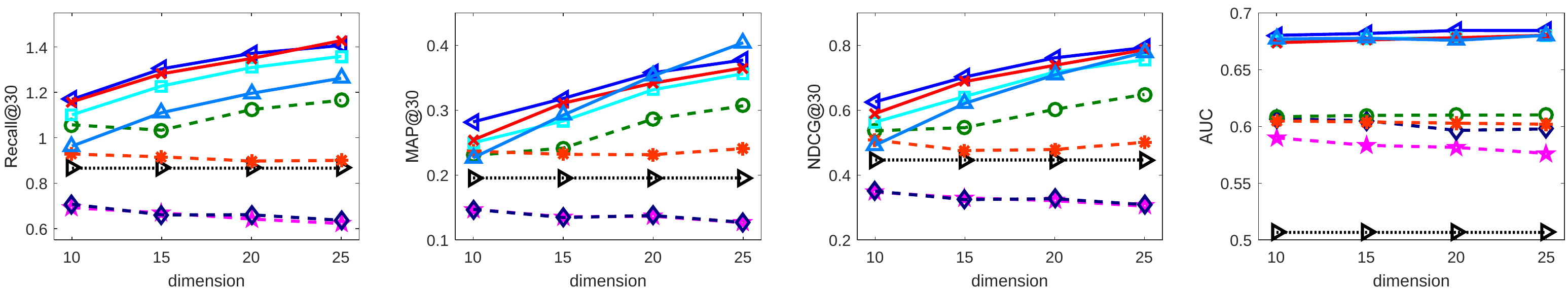}
    \label{figsub:taobao_Recall_varied_dimension_at_30}
    \end{minipage}
    }
    \subfigure[Amazon]{
    \begin{minipage}[b]{0.8\textwidth}
    \includegraphics[width=1\textwidth]{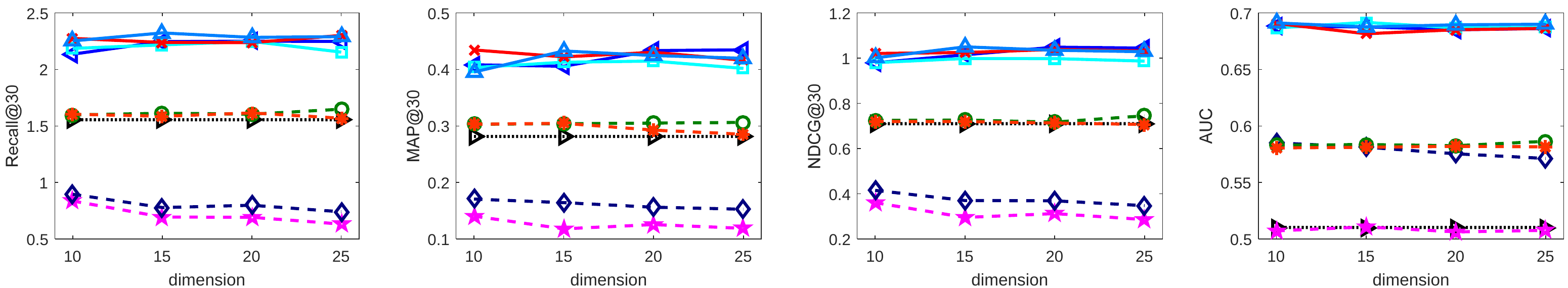}
    \label{figsub:amazon_Recall_varied_dimension_at_30}
    \end{minipage}
    }
    \caption{Recall@30, MAP@30, NDCG@30 and AUC performances on validation set with varied dimensions of latent feature $d=\left[10, 15, 20, 25\right]$.}
    \label{fig:dimension}
    \end{figure*}

    We design 3 combinations of input and 2 structures for the hidden state. There are 4 variants implemented as MV-RNN-Con., MV-RNN-Fus., MV-RNN-3mDAE-1U and MV-RNN-3mDAE-2U. The former 3 variants are built by the united structure, while the last one has the separate structure. The prefix `MV-RNN-' can be omitted, and the 4 variants can be abbreviated as \textbf{Con.}, \textbf{Fus.}, \textbf{3mDAE-1U} and \textbf{3mDAE-2U} respectively. The Con. has the highest dimension of hidden state ($\boldsymbol{h} \in \mathbb{R}^{3d}$), while the rest has the same dimension ($\boldsymbol{h} \in \mathbb{R}^{2d}$).
    Additionally, we need to initialize parameters $\Theta$ to the same range, e.g., uniform distribution $[-0.5, 0.5]$. The initial hidden state $\boldsymbol{h}^0$ of each sequence is always zero. The learning rate is fixed at $\alpha=0.1$ for all methods. Besides, the mini-batch size for training is set as 4 and users with similar lengths are grouped into one batch. This length-adjustment can greatly speed up training \cite{yang2016hierarchical}. Complete codes for all models are written by using Theano and are available on GitHub\footnote{https://github.com/cuiqiang1990/MV-RNN}.  All experimental results are also listed on this website.

\subsection{Optimization on Validation Set}
    \subsubsection{Regularization Parameter}
    \noindent
    The best parameters for regularization are listed in Table \ref{table:parameter_lambda}. They are chosen by the evaluations of all the metrics on validation set under the dimension $d=20$.

    In this optimization process, $\lambda_\Theta$ is firstly selected based on basic methods (BPR, GRU and LSTM), then $\lambda_\mathrm{ev}, r_\mathrm{a}$ are chosen by grid search. The ranges of these three parameters are $\lambda_\Theta, \lambda_\mathrm{ev} \in \left[0.001, 0.0001, 0.00001, 0.0\right]$ and $r_\mathrm{a} \in \left[0.001, 0.0001, 0.00001\right]$. With the reduction of data size from Taobao to Amazon, the best $\lambda_\Theta, \lambda_\mathrm{ev}, r_\mathrm{a}$ almost all get bigger.

    \begin{table*}[htbp]
      \centering\scriptsize
      \caption{Evaluation of different methods on the test set with the dimension of latent vector $d=20$. We generate top-20 and 30 items for each user. Because of the structure of concatenation, the hidden state dimension of Con. is much larger than the others.}
        \begin{tabular}{ccccccccc|cccccccc}
        \toprule
        & \multicolumn{8}{c}{Taobao} &\multicolumn{8}{c}{Amazon}\\
        \cmidrule(lr){2-9} \cmidrule(lr){10-17}
        \multirow{2}*{method} &\multirow{2}*{$p$}  &\multicolumn{3}{c}{@20 (\%)} &\multicolumn{3}{c}{@30 (\%)} & \multirow{2}*{AUC} &\multirow{2}*{$p$} &\multicolumn{3}{c}{@20 (\%)} &\multicolumn{3}{c}{@30 (\%)} &\multirow{2}*{AUC}\\
        \cmidrule(lr){3-5} \cmidrule(lr){6-8} \cmidrule(lr){11-13} \cmidrule(lr){14-16}
          & & Recall & MAP & NDCG & Recall & MAP & NDCG & & & Recall & MAP & NDCG & Recall & MAP & NDCG \\

        \midrule
          Random  & - & 0.004 & 0.001 & 0.002 & 0.006 & 0.001 & 0.003 & 0.500   & - & 0.083 & 0.016 & 0.040 & 0.137 & 0.018 & 0.056 & 0.500 \\
          POP     & - & 0.113 & 0.016 & 0.051 & 0.218 & 0.020 & 0.085 & 0.441   & - & 1.418 & 0.299 & 0.697 & 1.993 & 0.321 & 0.847 & 0.553 \\
          BPR     & - & 0.191 & 0.038 & 0.101 & 0.274 & 0.041 & 0.127 & 0.573   & - & 0.641 & 0.168 & 0.340 & 0.812 & 0.176 & 0.390 & 0.511 \\
          VBPR    & - & 0.196 & 0.042 & 0.106 & 0.283 & 0.045 & 0.131 & 0.577   & - & 0.700 & 0.181 & 0.368 & 0.922 & 0.190 & 0.423 & 0.584 \\
          LSTM    & - & 0.666 & 0.162 & 0.386 & 0.884 & 0.171 & 0.453 & 0.567   & - & 1.443 & 0.283 & 0.671 & 1.982 & 0.301 & 0.820 & 0.608 \\
          p-RNN   & - & 0.537 & 0.149 & 0.335 & 0.688 & 0.156 & 0.382 & 0.553   & - & 1.484 & 0.301 & 0.708 & 1.939 & 0.320 & 0.831 & 0.609 \\

        \midrule
          \textbf{Con.} & - & 0.863 & 0.212 & 0.502 & 1.164 & 0.224 & 0.592 & \textbf{0.690}   & - & 2.113 & 0.522 & 1.092 & 2.827 & 0.554 & 1.294 & \textbf{0.723} \\
          \textbf{Fus.} & - & 0.808 & 0.212 & 0.481 & 1.082 & 0.223 & 0.559 & \textbf{0.690}   & - & 2.157 & 0.508 & 1.096 & 2.867 & 0.538 & 1.285 & 0.722 \\

          \cmidrule(lr){2-2} \cmidrule(lr){10-10}
          \multirow{4}*{\textbf{3mDAE-1U}}
            & 0.0  & 0.849 & 0.213 & 0.499 & 1.140 & 0.225 & 0.586 & 0.680   & 0.0 & 2.190 & 0.517 & 1.116 & 2.869 & 0.549 & 1.309 & 0.722 \\
            & 0.2  & 0.802 & 0.205 & 0.472 & 1.075 & 0.216 & 0.555 & 0.687   & 0.1 & \textbf{2.243} & \textbf{0.541} & \textbf{1.149} & \textbf{2.995} & \textbf{0.570} & \textbf{1.352} & 0.722 \\
            & \textbf{0.3}  & \textbf{0.881} & 0.228 & \textbf{0.523} & \textbf{1.174} & 0.240 & \textbf{0.612} & 0.680   & 0.2 & 2.211 & 0.529 & 1.136 & 2.892 & 0.558 & 1.322 & 0.720 \\
            & 0.4  & 0.807 & 0.219 & 0.488 & 1.075 & 0.230 & 0.570 & 0.679   & 0.3 & 2.217 & 0.521 & 1.117 & 2.968 & 0.552 & 1.317 & 0.721 \\

          \cmidrule(lr){2-2} \cmidrule(lr){10-10}
          \multirow{4}*{\textbf{3mDAE-2U}}
            & 0.0  & 0.676 & 0.208  & 0.440 & 0.892 & 0.217 & 0.506 & 0.685  & 0.0 & 2.227 & 0.524 & 1.108 & 2.856 & 0.550 & 1.286 & 0.721 \\
            & 0.2  & 0.750 & 0.234 & 0.491 & 0.971 & 0.243 & 0.558 & 0.683   & 0.1 & 2.227 & 0.528 & 1.128 & 2.883 & 0.555 & 1.301 & 0.720 \\
            & 0.3  & 0.760 & 0.235 & 0.494 & 1.001 & 0.246 & 0.568 & 0.677   & 0.2 & 2.162 & 0.517 & 1.107 & 2.906 & 0.544 & 1.292 & 0.722 \\
            & 0.4  & 0.792 & \textbf{0.243} & 0.514 & 1.029 & \textbf{0.253} & 0.586 & 0.681   & 0.3 & 2.134 & 0.512 & 1.104 & 2.838 & 0.544 & 1.305 & 0.720 \\
        \bottomrule
        \end{tabular}
      \label{table:result}
    \end{table*}

    \subsubsection{Baseline Selection}
    \noindent
    Although several studies explore the difference between GRU and LSTM \cite{chung2014empirical,chung2015gated}, few people do comparisons for sequential recommendation. This part aims for completeness. Shown in Table \ref{table:gru_vs_lstm}, the result is the performance by using the best parameters obtained in Section 4.2.1.
    Please note that all values of Recall, MAP and NDCG in Tables \ref{table:gru_vs_lstm}, \ref{table:result}, \ref{table:result_sub_dataset}, \ref{table:missing_and_denoising}, \ref{table:cold_start} and Figure \ref{fig:dimension} are represented in percentage.

    Obviously, the performance of MV-RNN based on LSTM is better than that based on GRU in most cases, except the Con. and Fus. based on LSTM on Taobao. Although LSTM has more parameters, it also has the better model capacity. As a long as the model size is not significantly bigger, we should always consider the model with the best architecture. Therefore, in all the following experiments, we consider LSTM as the baseline instead of GRU and our MV-RNN is based on LSTM.

    \subsubsection{Dimension Analysis}
    \noindent
    The dimension analysis is investigated in Figure \ref{fig:dimension}. We illustrate the performances of top-$30$ and AUC on the validation set. The dimensions are set as $d=\left[10, 15, 20, 25\right]$.

    With the increasing of dimension, performances of top-30 metrics have similar trends with each other on both datasets.
    BPR and VBPR tend to get worse. They have similar trends as well as absolute values. It is difficult to tell the difference between VBPR and BPR on Recall, MAP, and NDCG, especially on Taobao. The p-RNN model is not sensitive to dimension.
    The LSTM and MV-RNN models obtain better performance with the increasing of dimension on Taobao, while they almost do not change with the dimension on Amazon.
    On the other hand, AUCs of all models are much stable with different dimensions. VBPR has obviously better performance than BPR on both datasets. The 4 variants of MV-RNN are nearly coincident with each other. The AUC is not sensitive to the dimension.

    Generally, it is obvious that LSTM is a very strong baseline. Apparently, our MV-RNN model is the best. The optimal dimension is chosen as $d=20$ and it is applied to other experiments.

    \begin{table}[tb]
      \centering\scriptsize
      \caption{Results of the controlled study in Section 4.3.2.}
        \begin{tabular}{ccrrrr}
        \toprule
        \multirow{2}*{dataset} &\multirow{2}*{method}  &\multicolumn{3}{c}{@30 (\%)} & \multirow{2}*{AUC} \\
        \cmidrule(lr){3-5}
          & & Recall & MAP & NDCG & \\

        \midrule
          \multirow{2}*{Taobao (10-core)}
            & LSTM          & 1.366 & 0.305 & 0.794 & 0.603\\
            & \textbf{Con.} & 1.635 & 0.365 & 0.946 & 0.689\\
        \midrule
          \multirow{2}*{Taobao (15-core)}
            & LSTM          & 2.343 & 0.752 & 1.742 & 0.591 \\
            & \textbf{Con.} & 2.801 & 0.868 & 2.040 & 0.678\\
        \midrule
          \multirow{2}*{Taobao (20-core)}
            & LSTM          & 4.681 & 13.795 & 16.651 & 0.536\\
            & \textbf{Con.} & 5.449 & 16.701 & 19.118 & 0.623\\
        \bottomrule
        \end{tabular}
      \label{table:result_sub_dataset}
    \end{table}

\subsection{Analysis of Experimental Results}
    \noindent
    Table \ref{table:result} illustrates all performances on two datasets with four evaluation metrics. Recall, MAP and NDCG focus on local performance, while AUC reflects global performance.

    \begin{table*}[htbp]
      \centering\scriptsize
      \caption{A setting called $missing$ is introduced and measured on an artificial test set, where some items' multi-modal features are missing (deleted). This setting aims to study the ability of MV-RNN to handle missing modalities.}
      \begin{tabular}{ccccccccc|cccccccc}
        \toprule
        & \multicolumn{8}{c}{$missing$ - Taobao}
        & \multicolumn{8}{c}{$missing$ - Amazon} \\
        \cmidrule(lr){2-9} \cmidrule(lr){10-17}
        \multirow{2}*{MV-RNN} & \multirow{2}*{$p$}
          & \multicolumn{3}{c}{@20 (\%)} & \multicolumn{3}{c}{@30 (\%)} & \multirow{2}*{AUC}
          & \multirow{2}*{$p$}
          & \multicolumn{3}{c}{@20 (\%)} & \multicolumn{3}{c}{@30 (\%)} & \multirow{2}*{AUC} \\
        \cmidrule(lr){3-5} \cmidrule(lr){6-8} \cmidrule(lr){11-13} \cmidrule(lr){14-16}
          & & Recall & MAP & NDCG & Recall & MAP & NDCG &
          & & Recall & MAP & NDCG & Recall & MAP & NDCG & \\

        \midrule
          \textbf{Con.} & - & 0.784 & 0.189 & 0.453 & 1.042 & 0.199 & 0.531 & 0.665   & - & 1.903 & 0.448 & 0.946 & 2.537 & 0.473 & 1.118 & 0.692 \\
          \textbf{Fus.} & - & 0.748 & 0.187 & 0.439 & 0.986 & 0.197 & 0.511 & 0.649   & - & 1.775 & 0.423 & 0.913 & 2.265 & 0.444 & 1.054 & 0.696 \\

          \cmidrule(lr){2-2} \cmidrule(lr){10-10}
          \multirow{4}*{\textbf{3mDAE-1U}}
            & 0.0 & 0.732 & 0.199 & 0.447 & 0.975 & 0.209 & 0.521 & 0.653  & 0.0 & 1.823 & 0.430 & 0.924 & 2.431 & 0.457 & 1.101 & 0.691 \\
            & 0.2 & 0.743 & 0.181 & 0.427 & 0.999 & 0.191 & 0.504 & \textbf{0.679}  & \textbf{0.1} & \textbf{2.059} & \textbf{0.49}1 & \textbf{1.040} & \textbf{2.696} & 0.517 & \textbf{1.217} & 0.703 \\
            & \textbf{0.3} & \textbf{0.832} & \textbf{0.216} & \textbf{0.496} & \textbf{1.102} & \textbf{0.228} & \textbf{0.578} & 0.671  & 0.2 & 2.003 & 0.488 & 1.028 & 2.561 & 0.510 & 1.176 & 0.702 \\
            & 0.4 & 0.746 & 0.191 & 0.440 & 1.000 & 0.202 & 0.517 & 0.666  & 0.3 & 2.001 & 0.470 & 0.995 & 2.645 & 0.498 & 1.171 & \textbf{0.705} \\

          \cmidrule(lr){2-2} \cmidrule(lr){10-10}
          \multirow{4}*{\textbf{3mDAE-2U}}
            & 0.0 & 0.605 & 0.181 & 0.388 & 0.791 & 0.188 & 0.444 & 0.652  & 0.0 & 1.779 & 0.392 & 0.864 & 2.414 & 0.414 & 1.042 & 0.688 \\
            & 0.2 & 0.643 & 0.180 & 0.400 & 0.851 & 0.189 & 0.464 & 0.673  & 0.1 & 1.964 & 0.452 & 0.979 & 2.624 & 0.479 & 1.163 & 0.704 \\
            & 0.3 & 0.676 & 0.194 & 0.424 & 0.897 & 0.204 & 0.492 & 0.670  & 0.2 & 1.858 & 0.482 & 0.986 & 2.476 & 0.506 & 1.145 & 0.702 \\
            & 0.4 & 0.701 & 0.201 & 0.441 & 0.920 & 0.210 & 0.508 & 0.674  & 0.3 & 1.827 & 0.496 & 1.003 & 2.435 & \textbf{0.522} & 1.171 & 0.704 \\
        \bottomrule
        \end{tabular}
      \label{table:missing_and_denoising}
    \end{table*}

    \subsubsection{Performance Comparison}
    \noindent
    From a global perspective, additional multi-modal information of items (e.g, image and text description) is indeed beneficial. VBPR beats BPR. MV-RNN outperforms LSTM model. Our MV-RNN can effectively model the additional information. For example, the Con. has almost more than 30\% and more than 40\% improvements over LSTM on Taobao and Amazon respectively with respect to Recall, MAP and NDCG. Its improvements of AUC over LSTM are both around 20\% on two datasets. As for the rest 3 variants which have hidden states of the same length, 3mDAE-1U performs best.
    In a perspective of statics and dynamics, although both trained by the BPR framework to maximize the difference of user's preferences towards positive and negative items, LSTM beats BPR by a large margin. The recurrent structure of LSTM can capture sequential information which is helpful for the recommendation.

    \noindent
    \textbf{3mDAE and Denoising}.
    In this part, we analyze the four variants of MV-RNN and focus on the 3mDAE.
    The Con. almost always beats the Fus. but not too much. The highest hidden state dimension of Con. improves its capacity. This phenomenon also shows that feature addition has no great damage to multi-modal modeling. Then, we embody the advantage of 3mDAE and introduce a training setting called $denoising$. It can help to learn more robust features and acquire the best performance.

    The denoising AE is first proposed for image classification on the MNIST database. 
    It can make features more robust and avoid learning the identity function by using corrupted input. Identity function means just mapping the original input to its copy, which happens in the encoding process in AE (e.g., $\mathbf{f} \rightarrow \boldsymbol{E}\mathbf{f}$).
    It is easy to obtain a denoising AE just by a stochastic corruption operation on input. The original corruption mechanism randomly sets some of an input feature to zero with probability $0\leqslant p<1$. While in our experiment, we make feature itself corrupted.

    This $denoising$ is conducted for 3mDAE. In this setting, we make some multi-modal data corrupted in the encoding process and still reconstruct both modalities in the decoding step. Training 3mDAE still requires all the data in Table \ref{tablesub:datasets_5_core}. The corruption levels are set as $p=\left[0.0, 0.2, 0.3, 0.4\right]$ and $p=\left[0.0, 0.1, 0.2, 0.3\right]$ for Taobao and Amazon respectively. If $p=0.0$, the input data in the encoding process is complete. The results are still obtained on the original test set where all items have all features.
    Results are shown in eight rows at the bottom of the Table \ref{table:result}.

    Obviously, performance can become better than the original ($p=0\%$) by $denoising$, especially the Recall, MAP and NDCG. More importantly, 3mDAE-1U performs best. It is able to be better than Con., although Con. has the highest hidden state dimension.
    When we randomly reset some features to zero in the encoding process, the noise in the whole input data is reduced. However, by reconstructing both modalities in the decoding step, the fusion feature of our 3mDAE can still keep the useful information in both modalities. Our 3mDAE can acquire more robust features. The best corruption levels for 3mDAE-1U/2U are $p=0.3/0.4$ and $p=0.1/0.1$ on two datasets respectively.

    The 3mDAE-1U/2U are a one-unit model with the united structure and a two-unit model with the separate structure respectively. In Table \ref{table:result}, the one-unit model outperforms the two-unit model. A united inner structure can better leverage the advantage of multi-view features. The separate structure may be not able to well model the connection between different views.

    \noindent
    \textbf{p-RNN vs. MV-RNN}. The session-based p-RNN model also incorporates additional features, but it is comparable to LSTM. If we carefully examine the results of p-RNN in its original paper \cite{hidasi2016parallel}, we find that most results of p-RNN are also close to the basic model (`ID only' in their paper). The reason is varied as p-RNN is substantially different from our MV-RNN. The first one is feature normalization. Multi-view features must be normalized to the same range, but only visual features are normalized in their work. Next, different from our strategy in Eq. \ref{preference}, p-RNN uses output weight matrix to compute the user's scores on items. This matrix improves the capacity of a model but increases the learning difficulty, especially for the modeling of visual and textual features.
    We experimented with using this matrix on our Con., but its performance is very close to that of LSTM.
    Then, different subnets within p-RNN are trained one by one, which can not well construct the connection among multi-view features.

    \begin{table*}[htbp]
      \centering\scriptsize
      \caption{Cold start performance on two datasets under the evaluation of Recall@30 and AUC with dimension of latent feature $d=20$.}
      \subtable[Numbers of items in each subset and each bin of the test set. Numbers of feedbacks are also counted.]{
        \begin{tabular}{ccrr|rrrrrrrrrr}
        \toprule
        \multirow{2}*{dataset}  & & \multicolumn{2}{c}{subsets of test set}  &\multicolumn{10}{c}{bins of test set} \\
        \cmidrule(lr){3-4}     \cmidrule(lr){5-14}
          & & $cold$-$start$ & $active$ & $[1, 2]$ & $[3, 4]$ & $[5, 6]$  & $[7, 8]$ & $[9, 10]$ & $[11, 12]$ & $[13, 14]$ & $[15, 16]$ & $[17, 18]$ & $[19, ]$ \\
        \midrule
        \multirowcell{2}{Taobao}
          & items       &  72,273 &   106,001   & 46,919 & 25,354 &  24,807 &  16,776 &  11,286 &  8,170 &  5,958 &  4,648 &  3,626 &    30,730  \\
          & feedbacks   & 152,623 & 2,918,957   & 64,363 & 88,260 & 135,031 & 124,920 & 106,703 & 93,649 & 80,135 & 71,916 & 63,380 & 2,243,223  \\
        \midrule
        \multirowcell{2}{Amazon}
          & items       & 12,399  &  2,826      & 8,970 & 3,429 & 1,422 & 525 & 340 & 184 & 98 & 64 & 49 & 144  \\
          & feedbacks   & 24,122  & 24,054      & 12,548 & 11,574 & 7,662 & 3,885 & 3,203 & 2,100 & 1,312 & 990 & 855 & 4,047  \\
        \bottomrule
        \label{tablesub:subset_interval}
        \end{tabular}}

      \subtable[Evaluation of cold start performance on Taobao. The interval is the accumulation of several bins.]{
        \begin{tabular}{cccccc|cccccccccccc}
        \toprule
        \multirow{2}*{eva.} &\multirow{2}*{method} &\multirow{2}*{p} &\multicolumn{3}{c}{subsets of test set (\%)} &\multicolumn{9}{c}{intervals of test set (\%)} \\
        \cmidrule(lr){4-6} \cmidrule(lr){7-16}
          & & & $cold$-$start$ & $active$ & $whole$ & $[1, 2]$ & $[1, 4]$ & $[1, 6]$  & $[1, 8]$ & $[1, 10]$ & $[1, 12]$ & $[1, 14]$ & $[1, 16]$ & $[1, 18]$ & $all$ \\
        \midrule
        \multirowcell{5}{Recall\\@30}
          & LSTM                & -   & 0.184 & 0.920 & 0.884 & 0.242 & 0.184 & 0.133 & 0.115 & 0.106 & 0.103 & 0.100 & 0.098 & 0.100 & 0.884 \\
          & \textbf{Con.}       & -   & 0.153 & 1.216 & 1.164 & 0.173 & 0.153 & 0.114 & 0.101 & 0.098 & 0.098 & 0.097 & 0.097 & 0.097 & 1.164 \\
          & \textbf{Fus.}       & -   & 0.144 & 1.131 & 1.082 & 0.174 & 0.144 & 0.109 & 0.105 & 0.103 & 0.103 & 0.106 & 0.105 & 0.109 & 1.082 \\
          & \textbf{3mDAE-1U}   & 0.3 & 0.269 & 1.221 & 1.174 & 0.362 & 0.269 & 0.195 & 0.171 & 0.165 & 0.160 & 0.157 & 0.157 & 0.158 & 1.174 \\
          & \textbf{3mDAE-2U}   & 0.4 & 0.621 & 1.050 & 1.029 & 0.839 & 0.621 & 0.437 & 0.378 & 0.354 & 0.340 & 0.333 & 0.328 & 0.324 & 1.029 \\
        \midrule
        \multirow{5}*{AUC}
          & LSTM                & -   & 0.608 & 0.565 & 0.567 & 0.657 & 0.608 & 0.519 & 0.487 & 0.473 & 0.467 & 0.463 & 0.462 & 0.462 & 0.567 \\
          & \textbf{Con.}       & -   & 0.659 & 0.691 & 0.690 & 0.681 & 0.659 & 0.631 & 0.623 & 0.620 & 0.620 & 0.619 & 0.621 & 0.621 & 0.690 \\
          & \textbf{Fus.}       & -   & 0.714 & 0.688 & 0.690 & 0.742 & 0.714 & 0.652 & 0.631 & 0.622 & 0.618 & 0.616 & 0.616 & 0.616 & 0.690 \\
          & \textbf{3mDAE-1U}   & 0.3 & 0.651 & 0.681 & 0.680 & 0.676 & 0.651 & 0.614 & 0.603 & 0.600 & 0.600 & 0.600 & 0.601 & 0.603 & 0.680 \\
          & \textbf{3mDAE-2U}   & 0.4 & 0.649 & 0.683 & 0.681 & 0.671 & 0.649 & 0.620 & 0.611 & 0.607 & 0.606 & 0.606 & 0.607 & 0.608 & 0.681 \\
        \bottomrule
        \label{tablesub:cold_start_taobao}
        \end{tabular}}

      \subtable[Evaluation of cold start performance on Amazon. The interval is the accumulation of several bins.]{
        \begin{tabular}{cccccc|cccccccccccc}
        \toprule
        \multirow{2}*{eva.} &\multirow{2}*{method} &\multirow{2}*{p} &\multicolumn{3}{c}{subsets of test set (\%)} &\multicolumn{9}{c}{intervals of test set (\%)} \\
        \cmidrule(lr){4-6} \cmidrule(lr){7-16}
          & & & $cold$-$start$ & $active$ & $whole$ & $[1, 2]$ & $[1, 4]$ & $[1, 6]$  & $[1, 8]$ & $[1, 10]$ & $[1, 12]$ & $[1, 14]$ & $[1, 16]$ & $[1, 18]$ & $all$ \\
        \midrule
        \multirowcell{8}{Recall\\@30}
          & LSTM                & -   & 0.000 & 3.970 & 1.982 & 0.000 & 0.000 & 0.000 & 0.003 & 0.033 & 0.034 & 0.066 & 0.074 & 0.165 & 1.982 \\
          & \textbf{Con.}       & -   & 0.398 & 5.263 & 2.827 & 0.215 & 0.398 & 0.538 & 0.676 & 0.826 & 0.996 & 1.135 & 1.192 & 1.398 & 2.827 \\
          & \textbf{Fus.}       & -   & 0.328 & 5.413 & 2.867 & 0.215 & 0.328 & 0.463 & 0.558 & 0.702 & 0.876 & 1.017 & 1.114 & 1.276 & 2.867 \\
          & \textbf{3mDAE-1U}   & 0.1 & 0.623 & 5.675 & 2.995 & 0.207 & 0.323 & 0.434 & 0.547 & 0.692 & 0.869 & 1.038 & 1.144 & 1.337 & 2.995 \\
          & \textbf{3mDAE-2U}   & 0.1 & 0.319 & 5.454 & 2.883 & 0.199 & 0.319 & 0.450 & 0.552 & 0.705 & 0.874 & 1.034 & 1.149 & 1.335 & 2.883 \\
        \midrule
        \multirow{8}*{AUC}
          & LSTM                & -   & 0.496 & 0.721 & 0.608 & 0.471 & 0.496 & 0.514 & 0.531 & 0.549 & 0.561 & 0.569 & 0.576 & 0.582 & 0.608 \\
          & \textbf{Con.}       & -   & 0.660 & 0.787 & 0.723 & 0.647 & 0.660 & 0.669 & 0.678 & 0.688 & 0.696 & 0.700 & 0.703 & 0.707 & 0.723 \\
          & \textbf{Fus.}       & -   & 0.667 & 0.777 & 0.722 & 0.654 & 0.667 & 0.676 & 0.683 & 0.691 & 0.697 & 0.701 & 0.704 & 0.707 & 0.722 \\
          & \textbf{3mDAE-1U}   & 0.1 & 0.656 & 0.788 & 0.722 & 0.640 & 0.656 & 0.667 & 0.677 & 0.687 & 0.694 & 0.698 & 0.702 & 0.705 & 0.722 \\
          & \textbf{3mDAE-2U}   & 0.1 & 0.658 & 0.783 & 0.720 & 0.645 & 0.658 & 0.666 & 0.675 & 0.685 & 0.692 & 0.697 & 0.700 & 0.703 & 0.720 \\
        \bottomrule
        \label{tablesub:cold_start_amazon}
        \end{tabular}}

      \label{table:cold_start}
    \end{table*}

    \subsubsection{A Controlled Study}
    \noindent
    In Table \ref{table:result}, the metrics (Recall, MAP and NDCG) seem to be low, especially on Taobao. Therefore, we conduct a controlled study to explore the factors that influence the metrics.

    Reducing the number of items (search space) may be helpful. We extract three sub-datasets from Taobao by increasing the filtering strategy as $[10, 15, 20]$-core. The statistics are shown in Table \ref{tablesub:datasets_bigger_core}. In this way, the search space is greatly reduced.
    Then, we perform experiments by using LSTM and Con.. Accordingly, we need to re-select the best parameters and the results are shown in Table \ref{table:result_sub_dataset}.
    With the increasing of $k$, the three metrics get bigger. Metrics of Taobao (20-core) are obviously bigger than that of the other datasts. This may be because the sparsity of Taobao (20-core) is clearly small. At the same time, our method Con. is always better than LSTM, which shows the effectiveness of our MV-RNN.
    Therefore, although the absolute values on Taobao are small, they are related to the dataset itself (e.g., sparsity).

    In summary, our MV-RNN model is better than the others. MV-RNN can well model multi-view features and achieves the best and stable performance in different situations. The $denoising$ of 3mDAE is a good setting to improve performance. Besides, special strategies used in p-RNN are not necessary for handling multi-view features. Feature concatenation is natural but very useful. A united structure with simultaneous training strategy is easy to use and is better than the separate subnets built for each view in p-RNN. These conclusions of joint learning are also confirmed by the previous works, like a multi-view model for cross-domain user modeling \cite{elkahky2015multi}.

\subsection{Analysis of Missing Modalities in Test Set}
    \noindent
    Multi-modal methods usually hold an assumption that all modalities are available. However, in practice, certain modality is often missing, like an item without the visual feature. In such case, our 3mDAE is theoretically better than the concatenation and fusion. To verify this, we introduce a setting of test set called $missing$. First, we artificially modify the test set. We set one-third of items without visual features, one-third without textual features, and the last one-third with all the multi-modal features. Then, the training procedure also applies the $denoising$, and the only difference between Sections 4.3 and 4.4 is that $missing$ here is evaluated on our artificial test set. The result is shown in Table \ref{table:missing_and_denoising}.

    Experimental results indicate that our 3mDAE is very promising for tackling missing modalities problem. Both 3mDAE-1U/2U perform very well and 3mDAE-1U is more successful. For example, 3mDAE-1U under $p=0.3$ increases by about 10 percent with respect to Con. on Recall, MAP and NDCG on Taobao. This improvement acquired by 3mDAE-1U under $p=0.1$ on Amazon is about 9 percent.
    Besides, 3mDAE-1U/2U also have some increases on AUC over Con. and Fus.. Our 3mDAE is greatly better than others in this $missing$ setting and it can effectively handle the items with missing modalities.

\subsection{Analysis of Cold Start}
\begin{figure*}[htbp]
    \centering
    \setlength{\abovecaptionskip}{0pt}
    \setlength{\belowcaptionskip}{-15pt}

    \includegraphics[width=0.6\textwidth]{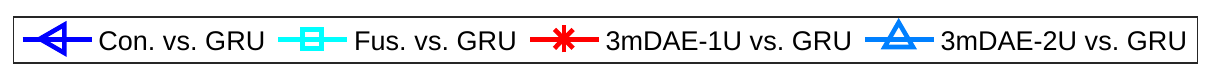}

    \subfigure[Taobao]{
    \begin{minipage}[b]{0.95\textwidth}  
    \includegraphics[width=1\textwidth]{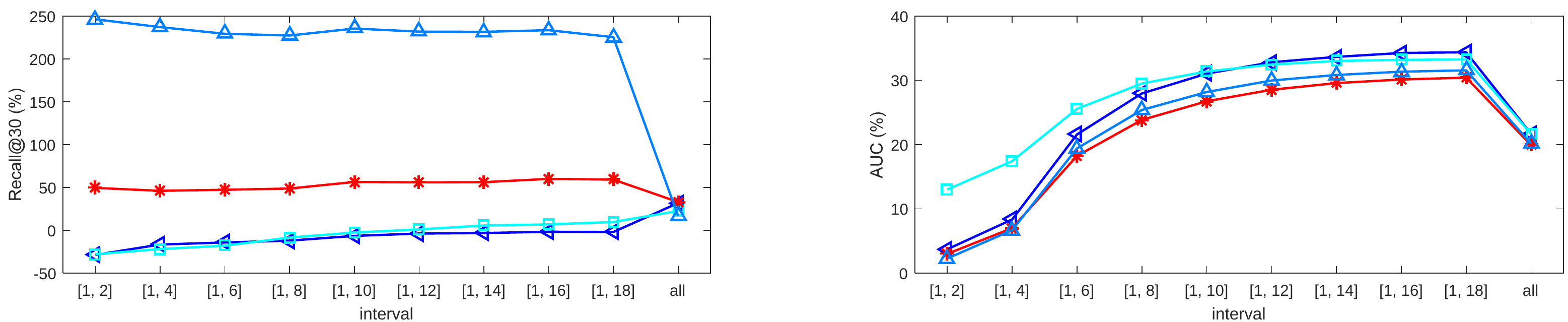}
    \label{figsub:taobao_cold_start}
    \end{minipage}
    }
    \subfigure[Amazon]{
    \begin{minipage}[b]{0.95\textwidth}
    \includegraphics[width=1\textwidth]{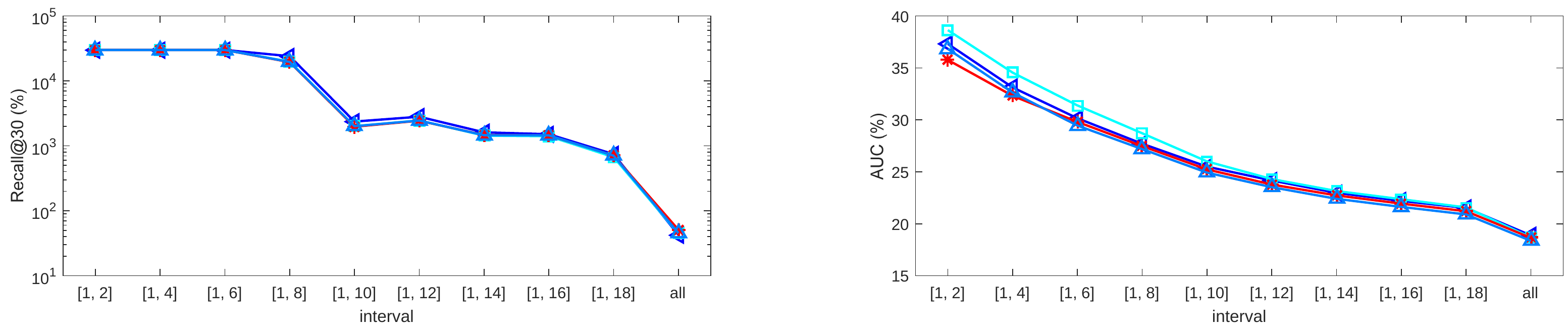}
    \label{figsub:amazon_cold_start}
    \end{minipage}
    }
    \caption{Based on the cold start performance in Table \ref{table:cold_start}, we calculate improvements (\%) on each interval. The best corruption levels $p$ for our 3mDAE-1U/2U is the same as in Table \ref{table:cold_start}, and we omit the $p$ in this figure.}
    \label{fig:cold_start_interval}
    \end{figure*}

    \noindent
    We investigate the performance of MV-RNN on cold start items in the test set. These items usually account for a large proportion and cold start is an intractable problem in practical recommender systems. Previous works like VBPR \cite{he2016vbpr} usually only consider cold start items and neglect the rest. While in our work, we expand this general setting because the rest items may produce a large volume of feedbacks. Two new experimental settings are designed, Recall@30 and AUC are applied to test the performance, as shown in Table \ref{table:cold_start}. Furthermore, we compute the improvement to analyze the effect of multi-modal information on cold start items.
    The improvements are shown in Table \ref{table:cold_start_subset} and Figure \ref{fig:cold_start_interval}.

    \begin{table*}[htbp]
      \centering\scriptsize
      \caption{Based on the cold start performance in Table \ref{table:cold_start}, we compute improvements (\%) on each subset. The best corruption levels $p$ for our 3mDAE-1U/2U is the same as in Table \ref{table:cold_start}, and we omit the $p$ in this table. The $cold$ refers to the $cold$-$start$.}
        \begin{tabular}{rrrrrrr|rrrrrr}
        \toprule
        \multirow{2}*{method$\qquad$} &\multicolumn{3}{c}{Taobao - Recall@30} &\multicolumn{3}{c}{Taobao - AUC}
        &\multicolumn{3}{c}{Amazon - Recall@30} &\multicolumn{3}{c}{Amazon - AUC} \\
        \cmidrule(lr){2-4} \cmidrule(lr){5-7} \cmidrule(lr){8-10} \cmidrule(lr){11-13}
          & $cold$ & $active$ & $whole$ & $cold$ & $active$ & $whole$
          & $cold$ & $active$ & $whole$ & $cold$ & $active$ & $whole$ \\
        \midrule
          \textbf{Con.} vs. LSTM      & -16.73 & 32.20 & 21.70 &  8.33 & 22.42 & 21.67   & 3$\times$10$^\text{4}$ & 32.57 & 42.62 & 33.12 & 9.07 & 18.88 \\
          \textbf{Fus.} vs. LSTM      & -22.05 & 22.87 & 22.41 & 17.41 & 21.88 & 21.64   & 3$\times$10$^\text{4}$ & 36.34 & 44.61 & 34.55 & 7.76 & 18.69 \\
          \textbf{3mDAE-1U} vs. LSTM  &  45.90 & 32.68 & 32.82 &  6.98 & 20.65 & 19.92   & 3$\times$10$^\text{4}$ & 42.93 & 51.10 & 32.31 & 9.23 & 18.65\\
          \textbf{3mDAE-2U} vs. LSTM  & 237.05 & 14.14 & 16.46 &  6.67 & 20.88 & 20.13   & 3$\times$10$^\text{4}$ & 37.38 & 45.45 & 32.63 & 8.53 & 18.36 \\
        \bottomrule
        \end{tabular}
      \label{table:cold_start_subset}
    \end{table*}

    \subsubsection{Subsets of Test Set}
    \noindent
    According to each item's support number in the test set, we divide items into three subsets: $cold$-$start$ ($\leqslant 4$), $active$ ($\geqslant 5$) and $whole$ (test set). Numbers of items of each subset are listed in Table \ref{tablesub:subset_interval}. The cold start items account for 40.5\% and 81.4\% on Taobao and Amazon respectively. 

    From the perspective of basic performance, as shown in Tables \ref{tablesub:cold_start_taobao} and \ref{tablesub:cold_start_amazon}, the best values are scattered in four variants. It is difficult to draw a consistent conclusion.

    As for the improvement shown in Table \ref{table:cold_start_subset}, most improvements on $cold$-$start$ are higher than those on $whole$, and are much higher than those on $active$.
    Comparatively, the basic model like LSTM has difficulty in predicting cold start items, while it is easier to obtain good performance on active items. Thus on the contrast, it is easy to design a model to substantially enhance the performance on $cold$-$start$, while it is more difficult to acquire obvious improvement on $active$.
    Under such situation, our MV-RNN still performs very well on $active$. For example, most improvements of MV-RNN are over 10\% on $active$.
    MV-RNN not only has a significant improvement on cold start items but also has a sufficient improvement on active items.

    In Table \ref{table:cold_start_subset}, there are some surprising improvements about Recall@30 on Amazon. We specify the improvement of MV-RNN over LSTM as $3\times10^\text{4}\%$, because the performance of LSTM on $cold$-$start$ is zero. This poor performance of LSTM can be explained from the perspective of probability.
    When we train a sequence, we practically apply LSTM to model a joint probability $p(x_1,\cdot\cdot\cdot,x_t)$,
    where $x_i$ represents an item.
    When we predict $n$ items in corresponding test sequence, we actually predict a conditional probability $p(x_{t+1},\cdot\cdot\cdot,x_{t+n} | x_1,\cdot\cdot\cdot,x_t)$. Because the 81.4\% cold start items and the corresponding 50.1\% feedbacks on Amazon result in limited interactions among users and items, both probabilities are very small. Therefore, it is very hard to make accurate recommendation under Recall@30 on Amazon.
    After we incorporate the additional content information, 4 variants of MV-RNN have performances of 0.398\%, 0.328\%, 0.623\% and 0.319\% respectively. The absolute values are small, but we obtain very large but reasonable improvements. This strange and extreme phenomenon exactly reflects the great power of additional content information and the powerful modeling capability of MV-RNN.

    \subsubsection{Intervals of Test Set}
    \noindent
    According to the support number of each item in the test set, we divide items into ten bins (e.g., $[1, 2]$, $[3, 4]$, $[5, 6]$). For example, bin $[1, 2]$ has the items that appear for 1 or 2 times. Numbers of items in each bin are listed in Table \ref{tablesub:subset_interval}.
    In order to alleviate the fluctuation of performance on each bin, performance is recorded on cumulative bins (e.g., $[1, 4]$) which are called intervals.


    When the bin number increases, performance becomes better, as seen from Tables \ref{tablesub:cold_start_taobao} and \ref{tablesub:cold_start_amazon}. That is because it is easier to predict frequent items. On Taobao, there is a strange phenomenon. Performance decreases first on a few bins in the front and then increases. As the decrement is not significant, we can still think the performance is growing. Then we mainly focus on the analysis of improvements. For better representation, improvements are illustrated by curves in Figure \ref{fig:cold_start_interval}.

    These growth curves do not always have the same change on two datasets. On Taobao, curves tend to be flat. On Amazon, as the bin has a larger proportion of cold start items (seeing from the right side of a figure to its left side), the improvement almost becomes larger.
    This indicates that multi-modal information is indeed beneficial to relieve cold start. In other words, when the cold start problem gets worse on small bins with a bigger proportion of cold start items, multi-modal information can significantly relieve this problem.
    Because cold start items have few interactions with users, directly related multi-modal information would effectively represent the item's characteristics and the user's interest.

    AUC is much more stable than Recall@30. We consider the difference of user's preferences towards positive and negative items in AUC, and the BPR training process exactly maximizes this difference.
    For Recall@30 curves, there is a large difference between Taobao and Amazon. These curves on Taobao are separate from each other, but they almost come together in the last interval $all$. Perhaps because of the small proportion of feedbacks on the interval $[1, 18]$ (27.0\%), there would be some fluctuations in the performance of each model.
    These curves on Amazon have an obvious increasing law when the bin number gets smaller.
    For AUC curves, the situation is much better. On Taobao, most improvements are stable. For example, improvements of MV-RNN are around 30\%. On Amazon, the smaller the bin number, the larger the improvement.

    These curves, especially those on Amazon, can greatly support the following conclusion. Multi-modal information can significantly relieve the item cold start problem. Besides, the worse the cold start, the more powerful the multi-modal information. 

    \begin{figure}[tb]
    \centering
    \setlength{\abovecaptionskip}{0pt}
    \setlength{\belowcaptionskip}{-5pt}
    \includegraphics[width=0.9\linewidth]{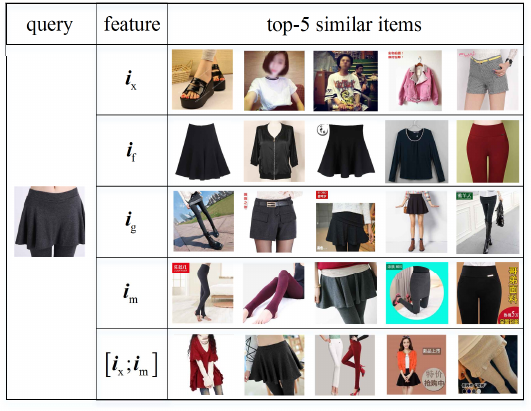}
    \caption{Visualization of similarity retrieval based on the Euclidean distance. Features are acquired by 3mDAE-1U under $p=0.3$ on Taobao.}
    \label{fig:similar}
    \end{figure}

    \subsubsection{Visualization of Learned Features}
    \noindent
    In this part, we make the visualization of learned features by similarity retrieval to investigate whether they are correlated or complementary. There are five different input features $\boldsymbol{i}_\mathrm{x}, \boldsymbol{i}_\mathrm{f}, \boldsymbol{i}_\mathrm{g}, \boldsymbol{i}_\mathrm{m}, [\boldsymbol{i}_\mathrm{x}; \boldsymbol{i}_\mathrm{m}]$ represented in Eqs. \ref{eq_multi_view_latent_ix}, \ref{eq_multi_view_if}, \ref{eq_multi_view_ig}, \ref{eq_multimodal_fusion_im}. Given a query item, we select top-5 most similar items based on the Euclidean distance for each kind of feature. The features are acquired by 3mDAE-1U under $p=0.3$ and the results are shown in Figure \ref{fig:similar}.

    Obviously, the similar items under different kinds of features vary greatly, and the multi-view (latent, visual, textual) features are complementary to each other.
    (1) For the latent feature $\boldsymbol{i}_\mathrm{x}$, the similar items are greatly different from each other as $\boldsymbol{i}_\mathrm{x}$ are just learned by the feedback. If the latent features of two items are similar, probably because they were both purchased by many people.
    (2) Whether it is item itself or the background in the image, the top-5 items based on the visual feature $\boldsymbol{i}_\mathrm{f}$ are very similar in appearance. However, the second and the forth items in this line obviously belong to other categories. The visual feature is powerful but can not reflect the intrinsic characteristics of items, like material of clothes.
    (3) On the other hand, the textual feature $\boldsymbol{i}_\mathrm{g}$ is acquired by the item description. It can truly reflect what the product is and can ignore the effect of the background in an image, but it is not intuitive to show the color, shape, etc.
    (4) The fusion feature $\boldsymbol{i}_\mathrm{m}$ is a combination of $\boldsymbol{i}_\mathrm{f}$ and $\boldsymbol{i}_\mathrm{g}$. It mainly integrates the external and intrinsic characteristics of the item, such as the style and material of clothes. However, such characteristics can not generate precise recommendation because there is no one-to-one match between each characteristic and each item.
    (5) The final item feature $[\boldsymbol{i}_\mathrm{x}; \boldsymbol{i}_\mathrm{m}]$ fuses $\boldsymbol{i}_\mathrm{x}, \boldsymbol{i}_\mathrm{f}, \boldsymbol{i}_\mathrm{g}$. It can fully reflect the characteristics of an item and help to understand the user's overall interest. In summary, multi-view features $\boldsymbol{i}_\mathrm{x}, \boldsymbol{i}_\mathrm{f}, \boldsymbol{i}_\mathrm{g}$ used in our work are complementary.

\section{Conclusion}
    \noindent
    In this work, we have proposed a novel multi-view recurrent model (MV-RNN) for sequential recommendation and alleviating the item cold start problem.
    First, we construct comprehensive item representation with latent, visual and textual features by three different combinations. A 3mDAE model is introduced to build the fusion feature based on visual and textual features.
    Then the user's interest is captured by the recurrent structure. We devise two types of inner structures to handle multi-view features.
    Next, we design a united objective function to combine the preference loss of BPR and the reconstruction loss of our 3mDAE.
    Experiments validate the state-of-the-art performance of MV-RNN. The fusion feature of 3mDAE helps to learn more robust features and tackle the missing modalities problem.
    Experiments confirm that a united inner structure can better leverage the advantage of multi-view features than a separate one. The multi-modal information like the image and text description could indeed significantly alleviate the item cold start problem.

    In the future, we would investigate the item detection and segmentation in images. The items in images often have a large proportion of unrelated background, especially in the Taobao dataset. We would like to obtain the more accurate item representation. These can motivate the model to improve performance.




\ifCLASSOPTIONcaptionsoff
  \newpage
\fi

\footnotesize
\bibliographystyle{IEEEtran}
\bibliography{bare_jrnl_compsoc_tkde_MV_GRU}

\begin{IEEEbiography}[{\includegraphics[width=1in,height=1.25in,clip,keepaspectratio]{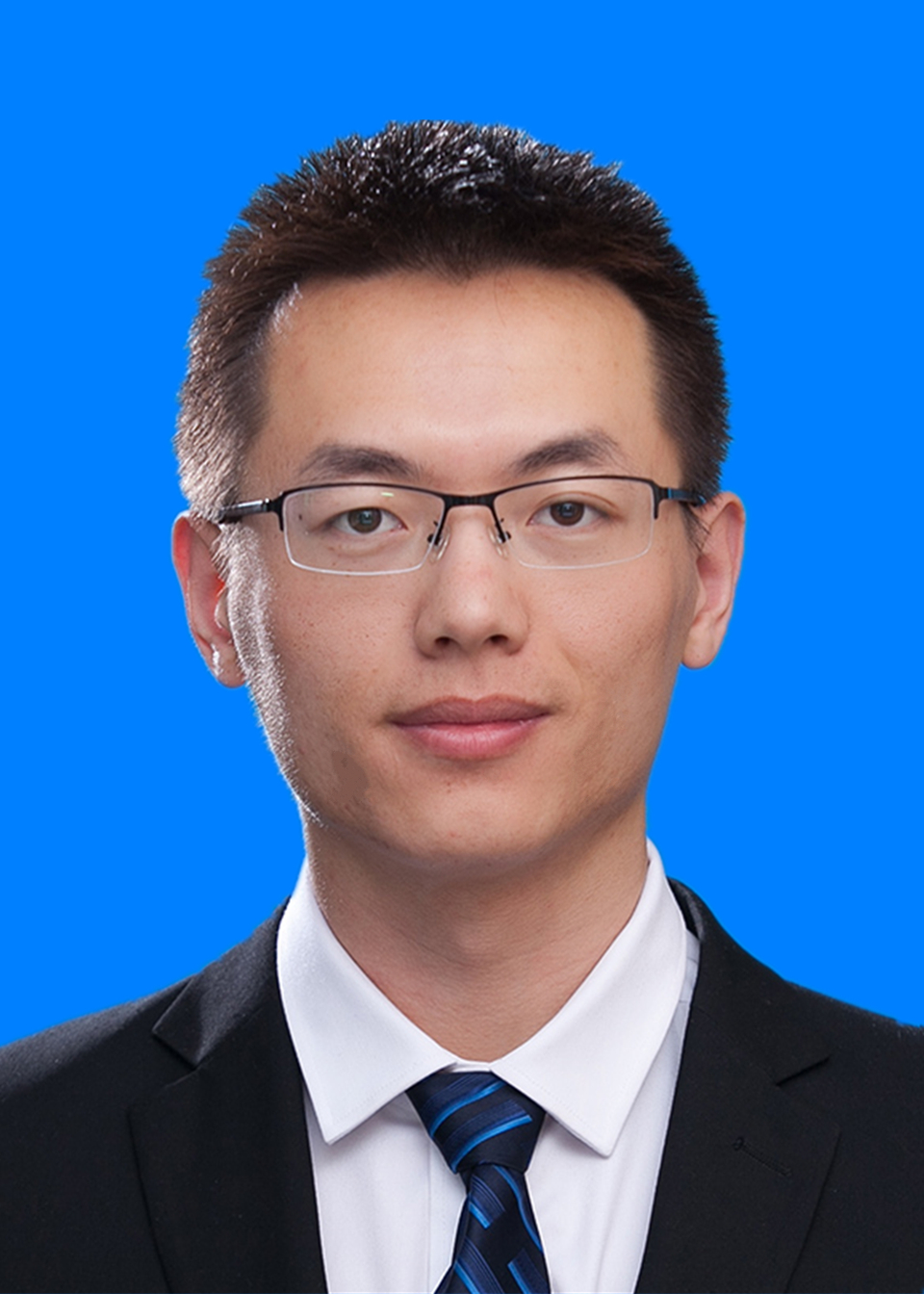}}]{Qiang Cui}
    received his B.S. degree from Shandong University, China, in 2013.
    He is currently working toward the Ph.D. degree in Center for Research on Intelligent Perception and Computing (CRIPAC) at National Laboratory of Pattern Recognition (NLPR), Institute of Automation, Chinese Academy of Sciences (CASIA), Beijing, China.
    His research interests include data mining, machine learning, recommender systems and information retrieval.
\end{IEEEbiography}

\begin{IEEEbiography}[{\includegraphics[width=1in,height=1.25in,clip,keepaspectratio]{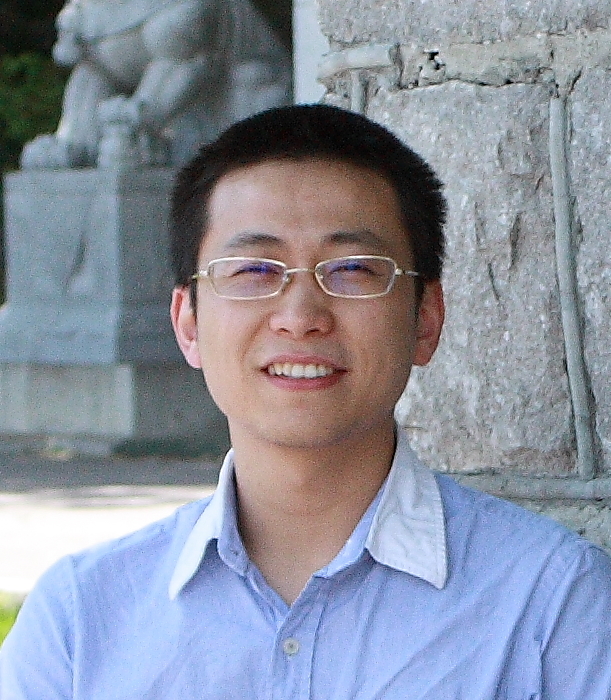}}]{Shu Wu}
    received his B.S. degree from Hunan University, China, in 2004, M.S. degree from Xiamen University, China, in 2007, and his Ph.D. degree from University of Sherbrooke, Quebec, Canada. He is an Associate Professor in Center for Research on Intelligent Perception and Computing (CRIPAC).
    He has published more than 20 papers in the areas of data mining and information retrieval at international journals and conferences, such as IEEE TKDE, IEEE THMS, AAAI, ICDM, SIGIR, and CIKM.
\end{IEEEbiography}

\begin{IEEEbiography}[{\includegraphics[width=1in,height=1.25in,clip,keepaspectratio]{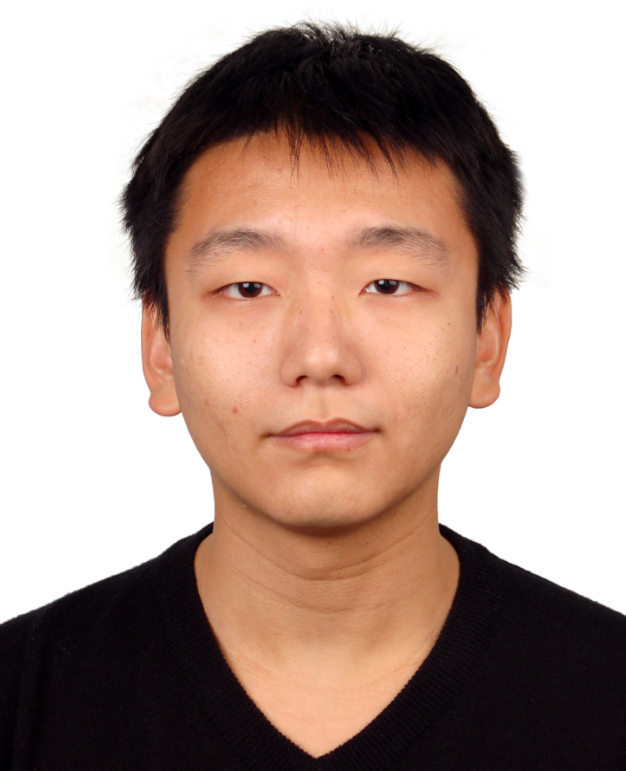}}]{Qiang Liu}
    received his B.S. degree in electronic science from Yanshan University, China, in 2013. He is currently working toward the Ph.D. degree in Center for Research on Intelligent Perception and Computing (CRIPAC).
    His research interests include machine learning, data mining, user modeling and information credibility evaluation. He has published several papers in the areas of data mining and information retrieval at international journals and conferences, such as IEEE TKDE, AAAI, SIGIR, CIKM, and ICDM.
\end{IEEEbiography}

\begin{IEEEbiography}[{\includegraphics[width=1in,height=1.25in,clip,keepaspectratio]{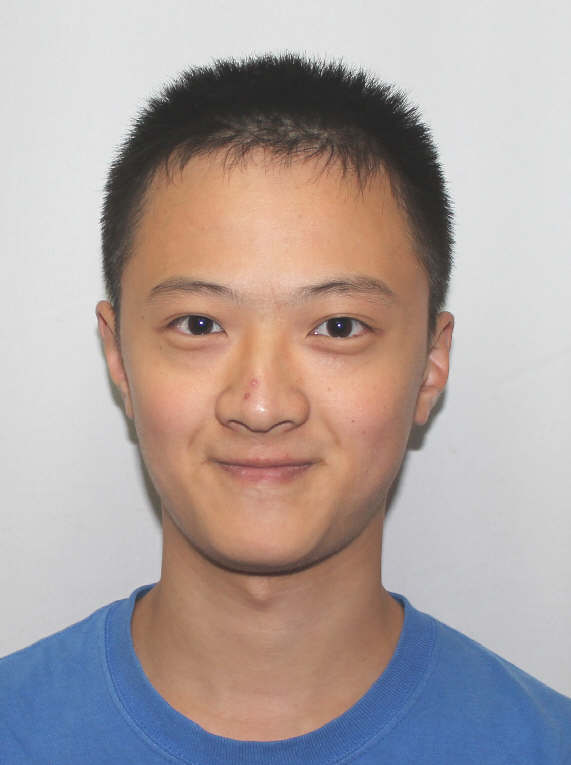}}]{Wen Zhong}
    received his B.S. degree in software engineering from Shandong University, China in 2015. He is currently working toward the M.S. degree in computer science at University of Southern California, United States. His research interests include machine learning and natural language processing.
\end{IEEEbiography}

\begin{IEEEbiography}[{\includegraphics[width=1in,height=1.25in,clip,keepaspectratio]{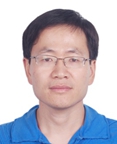}}]{Liang Wang}
    received both the BEng and MEng degrees from Anhui University in 1997 and 2000, respectively, and the Ph.D. degree from the Institute of Automation, Chinese Academy of Sciences (CASIA) in 2004.
    Currently, he is a full professor of the Hundred Talents Program at the National Lab of Pattern Recognition, CASIA. His major research interests include machine learning, pattern recognition, and computer vision. He has widely published in highly ranked international journals such as IEEE TPAMI and IEEE TIP and leading international conferences such as CVPR, ICCV, and ICDM. He is a senior member of the IEEE and an IAPR Fellow.
\end{IEEEbiography}




\end{document}